\PassOptionsToPackage{dvipsnames,table}{xcolor}

\documentclass[twocolumn]{aastex631}

\usepackage{threeparttable}
\usepackage{multirow}
\usepackage{enumitem}
\usepackage{natbib}
\usepackage{url}
\usepackage{amsmath,bm}
\usepackage{tikz}
\usepackage{float}
\usetikzlibrary{calc}
\usepackage{placeins}
\usepackage{diagbox}

\shorttitle{Cross-Resolution Stellar Parameter Estimation with DESI}
\shortauthors{Zhao et al.}

\begin{document}

\title{Generalization from Low- to Moderate-Resolution Spectra with Neural Networks for Stellar Parameter Estimation: A Case Study with DESI}

\correspondingauthor{Xiaosheng Zhao}
\email{xzhao113@jh.edu}

\author[0000-0002-8328-1447]{Xiaosheng Zhao}
\affiliation{Department of Physics \& Astronomy, The Johns Hopkins University, Baltimore, MD 21218, USA}

\author[0000-0001-5082-9536]{Yuan-Sen Ting}
\affiliation{Department of Astronomy, The Ohio State University, 140 West 18th Avenue, Columbus, OH 43210, USA}
\affiliation{Center for Cosmology and AstroParticle Physics (CCAPP), The Ohio State University, Columbus, OH 43210, USA}
\affiliation{Max-Planck-Institut f\"ur Astronomie, K\"onigstuhl 17, D-69117 Heidelberg, Germany}

\author[0000-0002-4013-1799]{Rosemary F.G. Wyse}
\affiliation{Department of Physics \& Astronomy, The Johns Hopkins University, Baltimore, MD 21218, USA}

\author[0000-0002-4108-3282]{Alexander S. Szalay}
\affiliation{Department of Physics \& Astronomy, The Johns Hopkins University, Baltimore, MD 21218, USA}
\affiliation{Department of Computer Science, The Johns Hopkins University, Baltimore, MD 21218, USA}

\author[0000-0003-3250-2876]{Yang Huang}
\affiliation{School of Astronomy and Space Science, University of Chinese Academy of Sciences, Beijing 100049, People's Republic of China}
\affiliation{National Astronomical Observatories, Chinese Academy of Sciences, Beijing 100012, People's Republic of China}

\author[0000-0001-7679-9478]{L\'aszl\'o Dobos}
\affiliation{Department of Physics \& Astronomy, The Johns Hopkins University, Baltimore, MD 21218, USA}
\affiliation{Department of Information Systems, E\"otv\"os Lor\'and University, Budapest 1117, Hungary}

\author[0000-0002-7034-4621]{Tam\'as Budav\'ari}
\affiliation{Department of Applied Mathematics \& Statistics, Johns Hopkins University, Baltimore, MD 21218, USA}
\affiliation{Department of Physics \& Astronomy, The Johns Hopkins University, Baltimore, MD 21218, USA}
\affiliation{Department of Computer Science, The Johns Hopkins University, Baltimore, MD 21218, USA}

\author{Viska Wei}
\affiliation{Department of Physics \& Astronomy, The Johns Hopkins University, Baltimore, MD 21218, USA}
\affiliation{Department of Computer Science, The Johns Hopkins University, Baltimore, MD 21218, USA}
\affiliation{Department of Applied Mathematics \& Statistics, Johns Hopkins University, Baltimore, MD 21218, USA}

\begin{abstract}

Cross-survey generalization is a critical challenge in stellar spectral analysis, particularly in cases such as transferring from low- to moderate-resolution surveys. We investigate this problem using pre-trained models, focusing on simple neural networks such as multilayer perceptrons (MLPs), with a case study transferring from LAMOST low-resolution spectra (LRS) to DESI medium-resolution spectra (MRS). Specifically, we pre-train MLPs on either LRS or their embeddings and fine-tune them for application to DESI stellar spectra. We compare MLPs trained directly on spectra with those trained on embeddings derived from transformer-based models (self-supervised foundation models pre-trained for multiple downstream tasks). We also evaluate different fine-tuning strategies, including residual-head fine-tuning, LoRA, and full fine-tuning. We find that MLPs pre-trained on LAMOST LRS achieve strong performance, even without fine-tuning, and that modest fine-tuning with DESI spectra further improves the results. For iron abundance, embeddings from a transformer-based model yield advantages in the metal-rich ([Fe/H] $>$ –1.0) regime, but underperform in the metal-poor regime compared to MLPs trained directly on LRS. We also show that the optimal fine-tuning strategy depends on the specific stellar parameter under consideration. These results highlight that simple pre-trained MLPs can provide competitive cross-survey generalization, while the role of spectral foundation models for cross-survey stellar parameter estimation requires further exploration.

\end{abstract}

\section{Introduction}
Large-scale spectroscopic surveys such as RAVE \citep{2006AJ....132.1645S}, SEGUE \citep{2010ApJ...714..663D}, APOGEE \citep{2017AJ....154...94M}, GALAH \citep{2015MNRAS.449.2604D}, LAMOST \citep{2012RAA....12..723Z}, and DESI \citep{2016arXiv161100036D} have revolutionized our understanding of the Milky Way by delivering millions of stellar spectra. A central challenge is how to to generalize parameter-estimation pipelines effectively, i.e., maintain accuracy and consistency across heterogeneous datasets from surveys that differ in their wavelength coverage, spectral resolution, signal-to-noise ratio, and in the stellar populations they target (e.g., ranges of mass, evolutionary phase, and elemental abundances). 

%Traditional template-fitting pipelines \citep{2009A&A...501.1269K, 2014IAUS..306..340W}, while robust within a given survey, often struggle to accommodate domain shifts between surveys and can be particularly sensitive to systematics and noise without careful treatment of differences in systematics and noise \citep{2019MNRAS.486.2075B}. 
Conventional machine-learning models \citep{2015ApJ...808...16N, 2017ApJ...849L...9T, 2019ApJ...879...69T,2019ApJS..245...34X, 2025ApJ...980...66R,2024ApJS..273...19Z}, although flexible and capable of learning to accommodate survey-specific systematics or noise, often suffer from domain mismatch when directly applied to a new survey. Their use is further limited by the small number of high-quality labels available in the target survey to train new models from scratch. Addressing this cross-survey generalization problem is therefore crucial for leveraging the full potential of current and upcoming spectroscopic datasets.

Recent advances in machine learning have highlighted the promise of foundation models\footnote{The ``Foundation'' model refer to a machine learning model that are pre-trained on a large base of datasets, with enough trainable parameters inside this model that are capable of capturing the full statistical relations in the datasets. Then this model is expected to be finetuned and do a good job for a specific domain or specific task, using a small-to-moderate amount of labeled datasets, assuming the successful pre-training in the previous stage.}, which have achieved success in natural language processing and computer vision \citep{NIPS2017_3f5ee243, 2018arXiv181004805D, radford2018improving, radford_language_2019, NEURIPS2020_1457c0d6}. In stellar spectroscopy, transformer-based models trained on large datasets can in principle provide general-purpose spectral representations that are transferable across tasks \citep{2024MNRAS.527.1494L,2024arXiv241104750K,2024arXiv241016081B,2024MNRAS.531.4990P,2024arXiv241108842R,2024arXiv240514930S,2024arXiv241221130Z,2025arXiv250315312E,2025arXiv250101070P,2025arXiv250701939Z,2025arXiv250720972Z}. However, the effectiveness of such spectral foundation models for cross-survey stellar parameter estimation remains a less-explored area. In parallel, simpler neural network architectures such as multilayer perceptrons (MLPs, \citealp{popescu2009multilayer}) may already capture much of the transferable information, if trained on sufficiently comprehensive datasets.

In this work, we investigate cross-survey generalization from the LAMOST Low-Resolution Spectra (LRS; $R \approx 1800$ over approximately 400-550~nm; \citealt{2012RAA....12..723Z}) to DESI\footnote{\url{https://www.desi.lbl.gov/spectrograph/}} spectra ($R \approx 2000$–3200 over 360--555~nm), and apply neural networks to extract intrinsic physical information, with fine-tuned models further learning to bridge the differences between the two surveys directly from the data. The LAMOST LRS, when cross-matched with high-resolution surveys such as APOGEE to get diverse labels from them, provides a rich training set. We pre-train MLPs on LAMOST LRS and evaluate their ability to generalize to DESI stellar parameter estimation, both in a \textit{zero-shot} setting (directly applying a pre-trained model to a new survey without fine-tuning) and in a \textit{few-shot} setting (using a small number of spectra from the new survey for fine-tuning). For comparison, we also explore transformer-based spectral embeddings as an example of a foundation-model approach, with and without contrastive alignment\footnote{Contrastive training encourages similar samples, particularly spectra originating from the same source, to have nearby embeddings and dissimilar samples to be well separated, thereby producing more robust representations.} to external surveys. 

Another focus of this study is the effectiveness of fine-tuning strategies. We compare full fine-tuning with lightweight adaptation methods such as residual-head fine-tuning and Low-Rank Adaptation (LoRA, \citealp{2021arXiv210609685H, 2025arXiv250720972Z}), and assess how different strategies affect generalization performance.

The goals of this paper are three fold: (1) to evaluate the performance of pre-trained MLPs for cross-survey stellar parameter estimation, compared with the existing DESI pipelines; (2) to evaluate whether MLPs pre-trained directly on spectra, once fine-tuned, achieve performance comparable to or exceeding that of MLPs fine-tuned on foundation-model spectral embeddings; and (3) to clarify the effectiveness of fine-tuning strategies in achieving robust transfer across surveys. 

The structure of this paper is as follows. In Section~\ref{sec:pretrained_mlp}, we describe the pre-trained MLPs and the spectral embeddings from transformer-based foundation models. In Section~\ref{sec:finetune_mlp}, we introduce the fine-tuning strategies. Section~\ref{sec:datasets} details the datasets used for pre-training, fine-tuning, and evaluation. Section~\ref{sec:results} presents results on DESI spectra, including zero-shot transfer and fine-tuning experiments. In Section~\ref{sec:discussion}, we present further discussion of foundation models, fine-tuning strategies, loss landscapes of different strategies, and the limitations of this work. Finally, we summarize our findings in Section~\ref{sec:summary}. Additional results are presented in the Appendices, including different ablation studies (Appendix~\ref{sec:ablation}), a saliency analysis of spectral features (Appendix~\ref{sec:sensitivity}), a summary of the dataset selection workflow adopted in the main analysis (Appendix~\ref{sec:dataset_cut}), and comparison with clean calibrated DESI SP subset (Appendix~\ref{sec:compare_clean}).

\begin{figure*}
    \begin{center}
    \includegraphics[scale=0.35,angle=0]{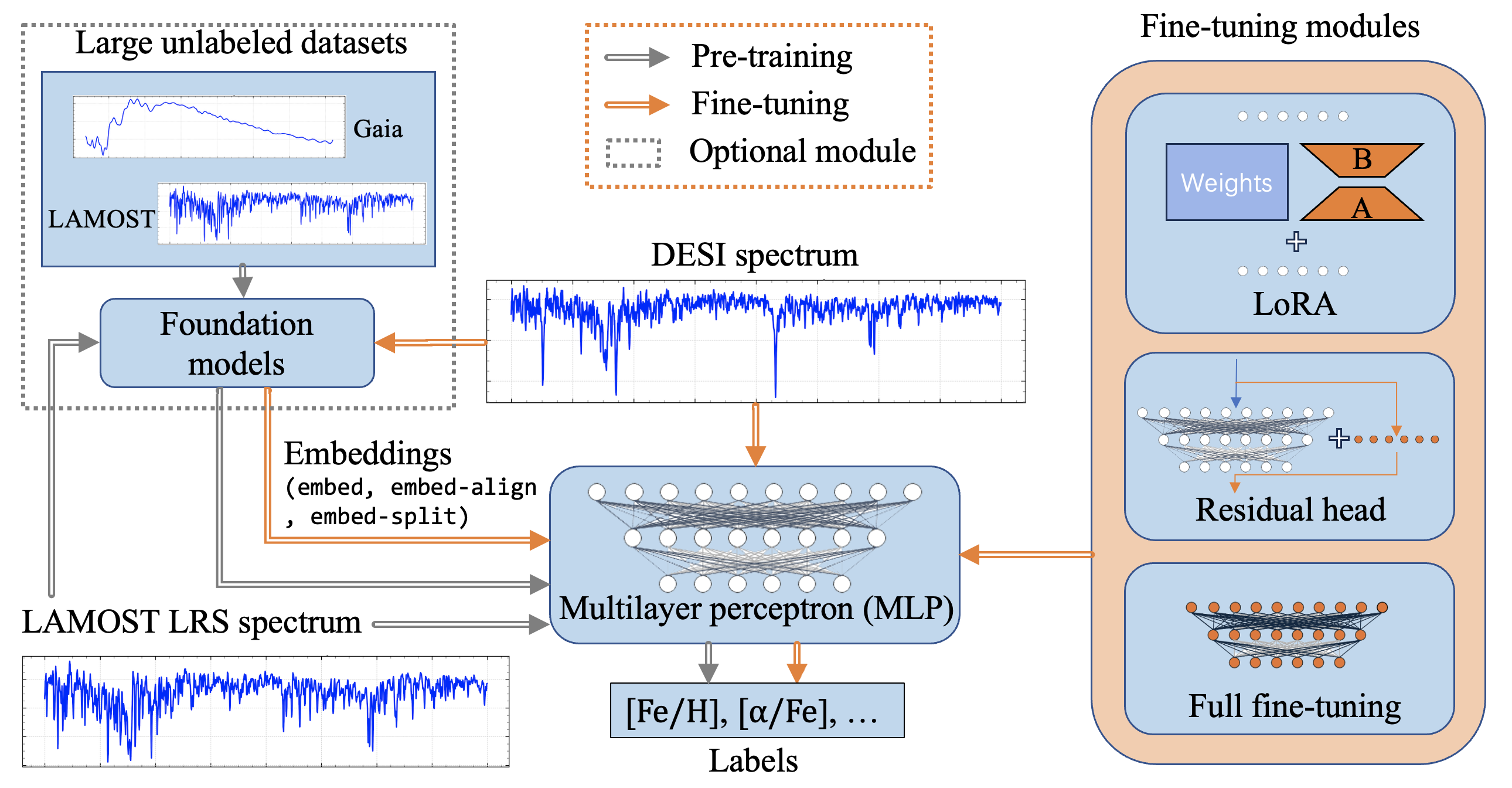} 
    \caption{Sketch of the pre-training and fine-tuning workflow. Pre-training uses normalized spectra or foundation-model spectral embeddings to predict labels: [Fe/H] from APOGEE ($>$ –2.0), supplemented at lower metallicities by PASTEL, SAGA, and other VMP/UMP datasets; [$\alpha$/Fe] from APOGEE. The foundation model is trained on large unlabeled datasets. Fine-tuning then adapts the network using fewer labeled spectra, either from spectra or spectral embeddings, with different fine-tuning modules applied (see Section~\ref{sec:finetune_mlp}).}
    \label{fig:sketch}
    \end{center}
\end{figure*}

\section{The MLP model for LAMOST LRS}
\label{sec:pretrained_mlp}

We pre-train simple multilayer perceptron (MLP, \citealp{popescu2009multilayer}) models on the LAMOST low-resolution spectra (LRS) dataset, which contains a moderate number ($\lesssim$ 100,000) of stellar spectra with labels obtained through cross-matching with high-resolution surveys such as APOGEE. One goal of this paper is to test whether such models, despite their simplicity, can provide robust cross-survey generalization.

The MLPs are designed to map spectral information to stellar parameters, with inputs given either as continuum-normalized spectra (hereafter referred to as \texttt{MLP-LRS}, or simply \texttt{lrs} in figures and tables) or as spectral embeddings (described in Section~\ref{sec:pretrained_foundation}), where the spectral fluxes are restricted to the 1,462 logarithmically spaced wavelength bins between 400–560 nm. The network consists of several fully connected layers with ReLU \citep{2019MNRAS.486.2075B} activations, followed by an output layer that predicts the target parameter. In this work we adopt the architecture $[\text{input\_dim}, 1024, 512, 64, 1]$, where $\text{input\_dim} = 1462$ when the input is continuum-normalized LRS or $768$ when the input is their embedding. The resulting model has $\sim 2.06$ million trainable parameters when applied directly to LRS fluxes.

We adopt the two pre-trained MLPs from \citet{2025arXiv250701939Z} to predict the logarithmic abundance ratios [Fe/H] and [$\alpha$/Fe]. The [Fe/H] model uses 90,106 LAMOST LRS with labels drawn from APOGEE DR17 \citep{2022ApJS..259...35A} for stars with [Fe/H] $>$ –2.0, and is supplemented at lower metallicities by the PASTEL and SAGA compilations \citep{2024ApJ...974..192H, 2008PASJ...60.1159S, 2016A&A...591A.118S}, the LAMOST/Subaru VMP sample \citep{2022ApJ...931..147L}, and UMP datasets \citep{2019MNRAS.484.2166S}. The [$\alpha$/Fe] model uses 85,400 spectra with APOGEE DR17 labels. In both cases the data are divided into training and validation sets with a 9:1 ratio. Training is performed with the AdamW \citep{loshchilov2018decoupled} optimizer (learning rate $1\times10^{-5}$, weight decay $1\times10^{-4}$, and batch size of 32). Each model is trained for 100 epochs, with the best checkpoint selected based on validation performance.

\subsection{Embeddings from spectral foundation models}
\label{sec:pretrained_foundation}

To assess whether spectral foundation models can provide improved transferable features, we also train MLPs on embeddings from the SpecCLIP\footnote{https://github.com/Xiaosheng-Zhao/SpecCLIP} \citep{2025arXiv250701939Z,zhao_2025_17824840} framework. SpecCLIP is designed to align stellar spectra obtained from surveys with different wavelength ranges, resolutions, and noise properties by producing a shared representation space, while also retaining survey-specific information useful for downstream tasks.

The framework consists of modality-specific encoders trained on large spectroscopic datasets. For LAMOST LRS (DR11\footnote{\url{https://www.lamost.org/dr11/}}), 966,082 high-quality spectra with $S/N_g > 50$ and $m_g < 15.8$ were selected. Continuum-normalized spectra are further standardized and tokenized into overlapping segments of length 20 (stride 10), yielding 146 tokens per spectrum. In addition, the logarithmic standard deviation measured before standardization is provided as a separate special token. A 6-layer transformer encoder maps a masked version of these tokens into a sequence of features, from which a 768-dimensional embeddings are produced and used for reconstruction. During training, six token segments are randomly masked and reconstructed, encouraging the encoder to capture global spectral structure. MLPs pre-trained on these embeddings are hereafter referred to as \texttt{embed} in Tables~\ref{tab:finetune_embed_finetune} and~\ref{tab:embed_ratio}.

For Gaia XP spectra \citep{Gaia1,Gaia2}, which consist of 343 flux points spanning 336--1021~nm, we train an autoencoder with fully connected layers on 1 million normalized spectra (each divided by its flux at 550 nm), of which roughly 80\% have matching LAMOST LRS counterparts, irrespective of signal-to-noise ratio. The XP and LRS encoders share the same embedding dimensionality (768) and parameter count ($\sim$42.7 million), allowing their outputs to be compared directly.

For the base SpecCLIP variant used in this work, contrastive training is performed using 820,568 paired LRS–XP spectra to align modalities. Projection heads map the encoder outputs into a shared latent space, with each head including a cross-attention mechanism that projects the modality-specific embeddings into another 768-dimensional vector (one per modality). The model is trained such that paired spectra from the same star are close in this space, while spectra from different stars are pushed apart. The resulting embeddings are designed to support a wide range of downstream tasks, including parameter estimation and similarity search. MLPs pre-trained on the LAMOST LRS embeddings from this alignment are hereafter referred to as \texttt{embed-align} in the tables.

Another SpecCLIP variant explicitly partitions the projected 768-dimensional embedding space into a shared 512-dimensional subspace and a modality-specific 256-dimensional subspace through separate projection networks. The shared subspaces of the two modalities are aligned via contrastive training, while for each modality both shared and modality-specific subspaces are used to reconstruct its input, thereby maximizing the mutual information \citep{2018arXiv180806670D} between the input and projected embeddings. MLPs pre-trained on the combined (shared + modality-specific) projected embeddings from the LAMOST LRS modality are hereafter referred to as \texttt{embed-split} in the tables.

\section{Fine-tuning strategies for MLPs}
\label{sec:finetune_mlp}

In this work, we explore fine-tuning strategies only for the MLPs. For MLPs pre-trained on different spectral embeddings, the components responsible for generating the embeddings are kept fixed. End-to-end fine-tuning of the full models may require substantially larger training sets, and we leave a systematic exploration of that direction to future work. The overall workflow of pre-training and fine-tuning is illustrated in Figure~\ref{fig:sketch}.

\subsection{LoRA fine-tuning}
\label{sec:lora}

To adapt pre-trained models to DESI spectra with limited labeled data, we explore parameter-efficient fine-tuning using Low-Rank Adaptation (LoRA, \citealp{2021arXiv210609685H, 2025arXiv250720972Z}). Instead of updating all network parameters, LoRA inserts small low-rank matrices into selected linear layers. This allows the original pre-trained weights to remain frozen while a modest number of additional parameters capture survey-specific adjustments.

Formally, a weight update $\Delta W \in \mathbb{R}^{m \times n}$ is decomposed as $\Delta W = A B$, where $A \in \mathbb{R}^{m \times r}$ and $B \in \mathbb{R}^{r \times n}$ with rank $r \ll \min(m,n)$. This greatly reduces the number of trainable parameters while maintaining expressive capacity. In our implementation, LoRA modules are added to all linear layers of the MLPs. We experiment with LoRA configurations using a rank of 128 and a scaling factor of $\alpha=256$, introducing $\sim5.89 \times 10^5$ additional parameters (about 22\% of the MLP size).

\subsection{Residual-head fine-tuning}
\label{sec:residual_head}

As an alternative, we use a residual-head strategy in which a small MLP, operating in parallel with the frozen pre-trained MLP, is added to correct its outputs. The residual head has the architecture $[\text{input\_dim}, 384, 1]$, with ReLU activation. This introduces $\sim5.62 \times 10^5$ trainable parameters ($\sim$21\% of the base MLP size). By restricting adaptation to the residual head, this method balances flexibility with a reduced risk of overfitting.

\subsection{Full fine-tuning}
\label{sec:full_finetune}

For completeness, we also apply full fine-tuning, updating all (100\%) parameters of the pre-trained MLP. This strategy provides maximal adaptation flexibility but carries an increased risk of overfitting when training data are limited.

\subsection{Fine-tuning details}
\label{sec:training_details}

All fine-tuning is performed with the AdamW \citep{2017arXiv171105101L} optimizer using a learning rate of $1\times10^{-5}$, weight decay of $1\times10^{-4}$, and a batch size of 32. Models are trained for up to 200 epochs, and the best checkpoint is chosen based on the validation performance. Experiments are run on a single NVIDIA V100 GPU; training time depends on the fine-tuning strategy and the size of the fine-tuning dataset; full fine-tuning with the maximum sample size (about 2,000), including testing, takes about 20~s.

\section{Datasets}
\label{sec:datasets}

We use spectra from the DESI Early Data Release (EDR; \citealp{2024MNRAS.533.1012K}) and Data Release 1 (DR1; \citealp{2025arXiv250514787K}), accessed via the SPectra Analysis \& Retrievable Catalog Lab (SPARCL; \citealp{2024arXiv240105576J, 9347681}) at NOIRLab's Astro Data Lab \citep{10.1117/12.2057445, NIKUTTA2020100411}. DESI DR1 is used for fine-tuning (or for training from scratch as a baseline) and for testing the MLP models, owing to its substantially larger cross-match with APOGEE compared to DESI EDR. DESI EDR is used exclusively to evaluate the resulting [$\alpha$/Fe]–[Fe/H] distribution as an initial demonstration, as this dataset has also been examined by independent groups \citep{2024ApJS..273...19Z}. All spectra are re-gridded to the same wavelength bins as the LAMOST LRS using linear interpolation and continuum-normalized in the same manner (see Section~\ref{sec:pretrained_foundation}). A summary of the sample selection cuts is provided in Appendix~\ref{sec:dataset_cut}. In the main text, we select samples to retain most of the cross-matched sources, with additional conditions outlined in Appendix~\ref{sec:compare_clean}. 

For DESI DR1, we first cross-match the catalog with APOGEE DR17 using the Gaia DR3 \texttt{source\_id} provided in the DESI catalog to obtain high-quality labels. From the APOGEE side, we require \texttt{STARFLAG=0}, \texttt{FE\_H\_FLAG=0}, \texttt{MG\_FE\_FLAG=0}, and a median SNR per pixel (at \texttt{apStar} sampling) greater than 40. From the DESI side, we additionally require \texttt{RVS\_WARN=0} and, when available, \texttt{PRIMARY=True} (a small number of entries lack this flag). This yields 8,104 common stars. For fine-tuning or training from scratch, we select a subsample\footnote{This cut avoids low-temperature stars, for which APOGEE ASPCAP \citep{2016AJ....151..144G} abundances are reported to be less reliable (see also the SDSS DR17 abundance documentation: \url{https://www.sdss4.org/dr17/irspec/abundances/}). In the testing stage, we loosen this constraint to $T_{\rm eff}>4000$~K to enable broader comparison.} of stars with $T_{\rm eff}>4500$~K, consisting of all 469 stars with [Fe/H] $<-0.7$ and a random subset of 1,600 stars from the more metal-rich population. This produces a fine-tuning/training set of 2,069 stars. From the remaining 6,035 stars, we apply a cut of $T_{\rm eff}>4000$~K, resulting in 5,539 stars used exclusively for testing.

From this same cross-matched sample, we also set aside the full set of 7,608 stars with $T_{\rm eff}>4000$~K to construct the [$\alpha$/Fe]–[Fe/H] diagram using APOGEE labels, providing a direct comparison to MLP-based predictions from DESI EDR.

For DESI EDR, we apply the following quality cuts: \texttt{RVS\_WARN=0}, \texttt{PRIMARY=True}, \texttt{RR\_SPECTYPE=STAR}, and \texttt{MEDIAN\_COADD\_SNR\_B$>$20}. This yields 134137 stars with unique \texttt{TARGETID}. We further restrict the sample to stars with $T_{\rm eff}>4000$~K (based on DESI SP pipeline estimates), remove sources with missing [Fe/H] or [$\alpha$/Fe] values, and require successful flux normalization. The final EDR sample contains 126,253 stars, which we use to evaluate the [$\alpha$/Fe]–[Fe/H] distribution from the MLP predictions. Because only a small fraction of these spectra have cross-matched high-quality labels, this sample is used primarily for visual inspection and qualitative comparison with APOGEE DR17 labels cross-matched to DESI DR1.

\begin{figure*}
    \begin{center}
    \includegraphics[scale=0.25,angle=0]{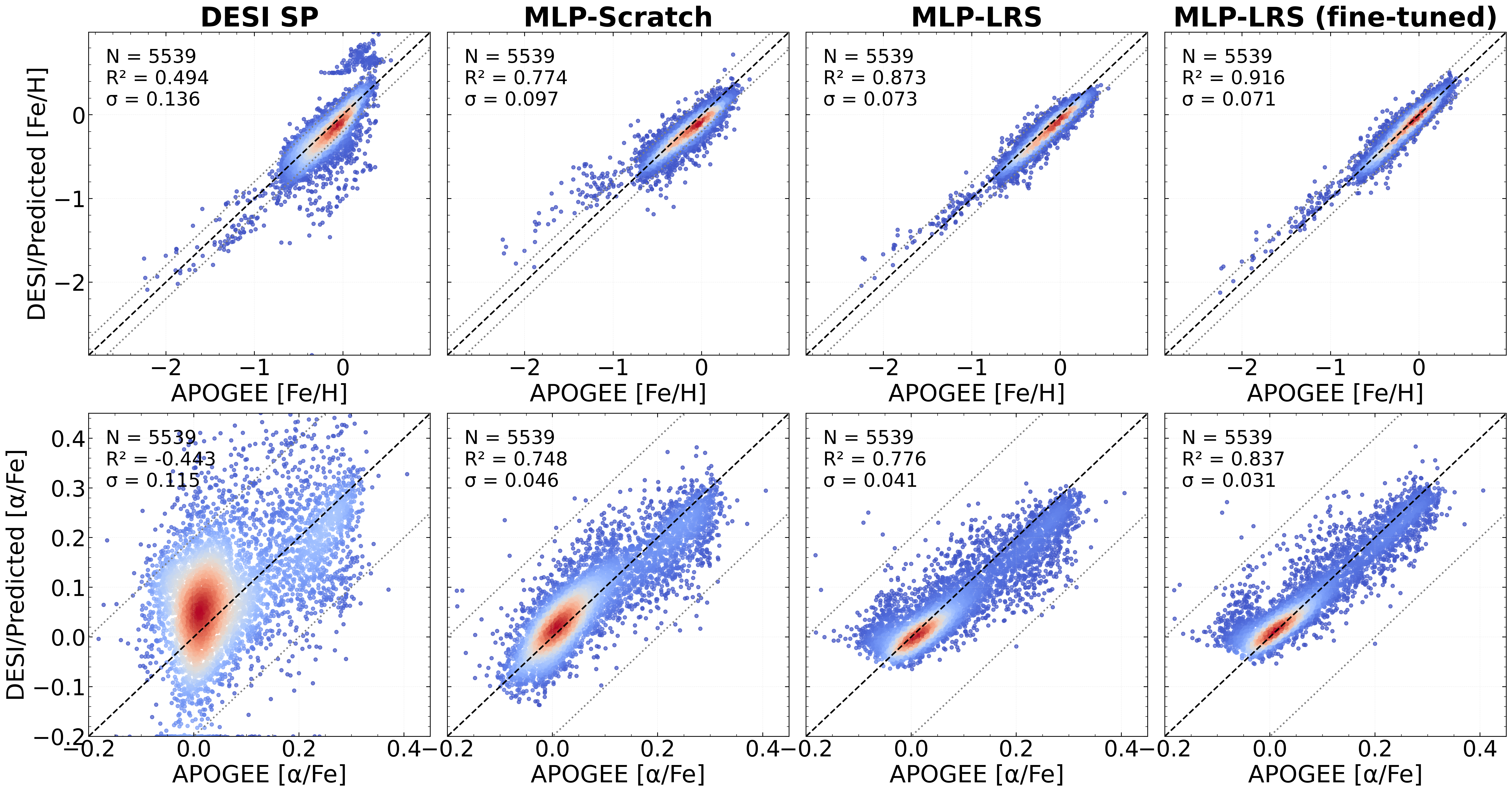} 
    \caption{Comparison of [Fe/H] and [$\alpha$/Fe] estimates between the DESI SP pipeline and MLP-based models, referenced against APOGEE DR17 labels. From left to right: \texttt{DESI SP}--DESI SP pipeline, \texttt{MLP-Scratch}--MLP trained from scratch on the same number of DESI spectra used for fine-tuning, \texttt{MLP-LRS}--MLP pre-trained on LAMOST LRS (zero-shot application), and \texttt{MLP-LRS (fine-tuned)}--MLP pre-trained on LAMOST LRS and fine-tuned with DESI spectra. The first row shows [Fe/H] results; the second row shows [$\alpha$/Fe]. Legends list the number of test stars (N), coefficient of determination ($\mathrm{R^2}$), and the robustly estimated standard deviation of the residuals ($\sigma$, computed after 3$\sigma$ clipping with \texttt{sigma\_clip} in \texttt{astropy}). All sources are restricted to $T_{\rm eff}>4000$~K. The dashed black line indicates the ideal 1:1 relation, and the gray dotted lines mark ±0.2 dex deviations. The fine-tuning set consists of 2,069 stars. A comparison using the clean, calibrated DESI SP subset is provided in Appendix~\ref{sec:compare_clean}.}
    \label{fig:compare_pretrained_desi}
    \end{center}
\end{figure*}

\begin{figure*}
    \begin{center}
    \includegraphics[scale=0.23,angle=0]{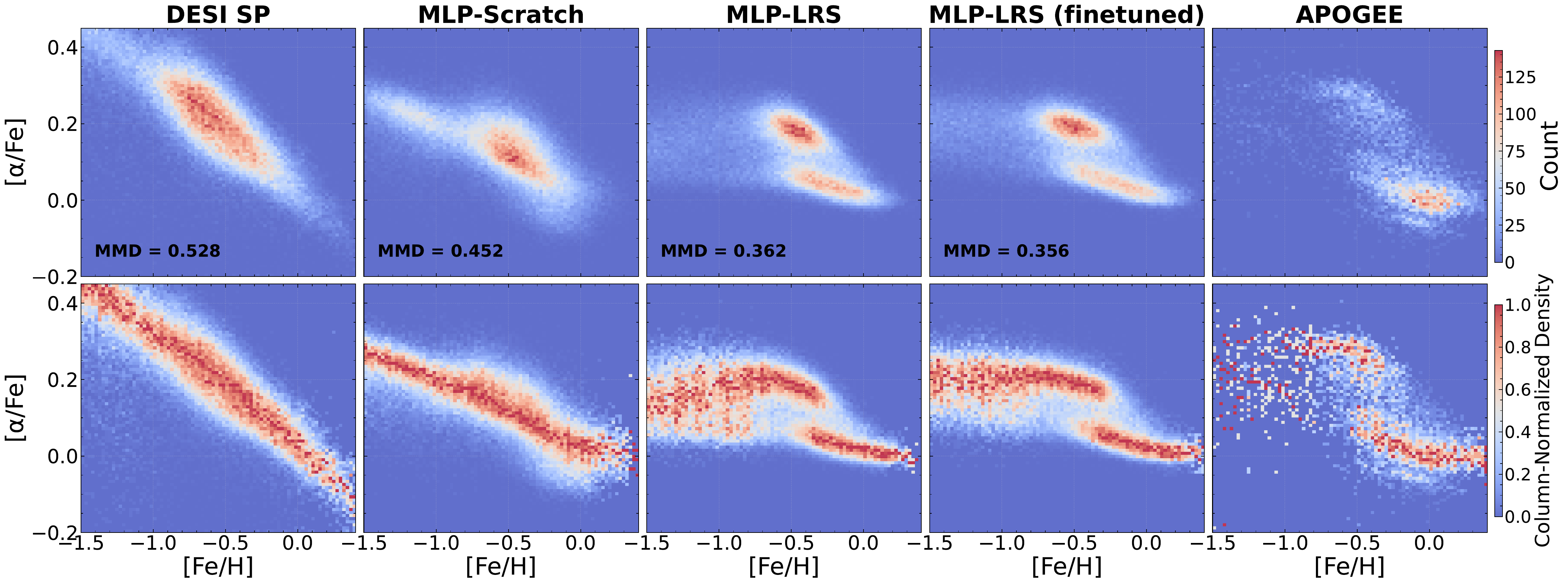} 
    \caption{[$\alpha$/Fe]–[Fe/H] diagrams for DESI EDR and APOGEE reference sample. From left to right: (1) \texttt{DESI SP}: DESI SP pipeline, (2) \texttt{MLP-Scratch}: MLP trained from scratch on the same number of DESI spectra used for fine-tuning, (3) \texttt{MLP-LRS}: MLP pre-trained on LAMOST LRS (zero-shot application), (4) \texttt{MLP-LRS (fine-tuned)}: MLP pre-trained on LAMOST LRS and fine-tuned with DESI spectra, and (5) \texttt{APOGEE}: DESI DR1–APOGEE DR17 cross-matched sample (7,608 stars). The first row shows raw counts, and the second row shows column-normalized densities, where each [Fe/H] column is divided by its maximum count. The legend reports the Maximum Mean Discrepancy (MMD) between each sample and the APOGEE reference.}
    \label{fig:compare_pretrained_desi_afe_feh}
    \end{center}
\end{figure*}

\begin{figure}
    \begin{center}
    \includegraphics[width=\columnwidth]{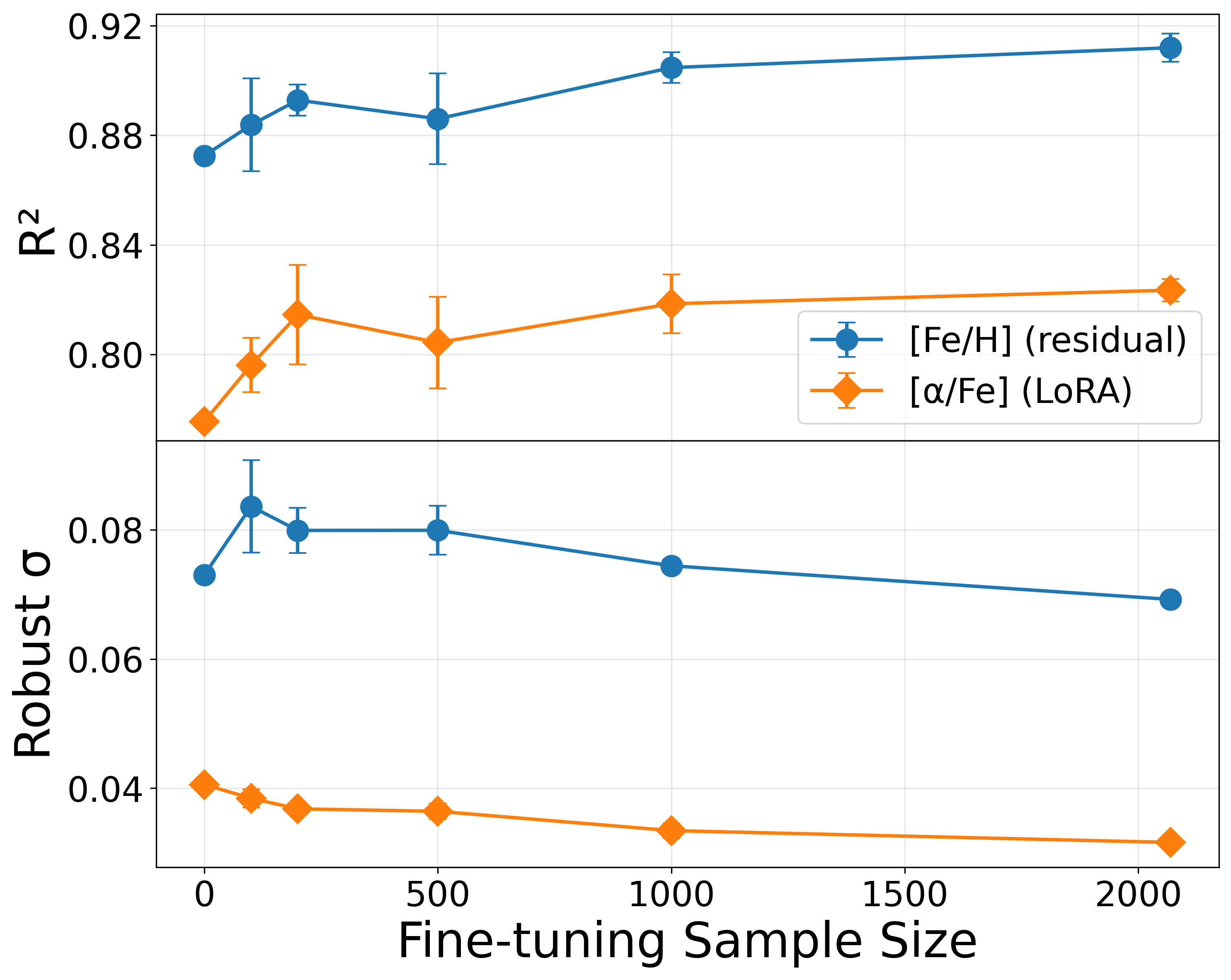} 
    \caption{Effect of fine-tuning sample size (0, 100, 200, 500, 1000, and 2069, where 0 denotes zero-shot performance), evaluated on the testing sample. Shown are $R^2$ and the robustly estimated standard deviation of the residuals ($\sigma$) for [Fe/H] and [$\alpha$/Fe], obtained after fine-tuning (residual-head fine-tuning for [Fe/H] and LoRA fine-tuning for [$\alpha$/Fe]). For both metrics, we plot the mean and 1$\sigma$ error bars from five independent runs. The scatter plots for the different sample sizes are shown in Figure~\ref{fig:sample_size_one2one} and \ref{fig:sample_size_2d}.}
    \label{fig:sample_effects}
    \end{center}
\end{figure}

\section{Results}
\label{sec:results}

\subsection{Testing the MLP model with DESI spectra}
\label{sec:test_mlp}

We first evaluate the zero-shot performance of the MLP models pre-trained on LAMOST LRS, applying them directly to DESI DR1 and EDR spectra. We focus on iron abundance ([Fe/H]) and $\alpha$-elemental abundance ([$\alpha$/Fe]), and compare the results with the DESI stellar parameter pipeline and with APOGEE labels for cross-matched stars. As an additional baseline, we also train MLPs from scratch using the same number of DESI spectra as in the fine-tuning experiments.

\paragraph{\textnormal{[$\alpha$/Fe]} and \textnormal{[Fe/H]} abundances}
Figure~\ref{fig:compare_pretrained_desi} shows that the zero-shot performance of the pre-trained MLP (\texttt{MLP-LRS}) provides improved agreement with APOGEE reference labels compared to both the DESI SP pipeline\footnote{Note that the SP pipeline uses \texttt{FERRE} \citep{2023ascl.soft01016A}, a template-fitting method, in both the EDR and DR1 releases. The RVS pipeline performs comparably to, or worse than, the SP pipeline, and is therefore not shown here; see \citet{2025arXiv250514787K}. Our analysis includes the full metallicity range available in the sample, including the metal-rich regime that is sparsely represented in some previously published comparisons.}(\texttt{DESI SP}) and models trained from scratch (\texttt{MLP-Scratch}) on DESI data, underscoring the value of pre-training even for simple architectures. We note that for [$\alpha$/Fe], only Ca and Mg in the DESI estimates are considered reliable \citep{2025arXiv250514787K}, and the increased scatter around [$\alpha$/Fe] $\lesssim 0$ is largely due to low-temperature stars with $T_{\rm eff}$ around $4000$~K. The figure legend reports the coefficient of determination and robust scatter (standard deviation of the residuals) for each method. For further reference, the comparison using the clean, calibrated DESI SP subset is provided in Appendix~\ref{sec:compare_clean}.

\paragraph{\textnormal{[$\alpha$/Fe]}–\textnormal{[Fe/H]} diagram}  
We also construct the [$\alpha$/Fe]–[Fe/H] distribution using DESI EDR spectra (Figure~\ref{fig:compare_pretrained_desi_afe_feh}, adopting axis limits of [Fe/H] $\in [-1.5,0.4]$ and [$\alpha$/Fe] $\in [-0.2,0.45]$ for visualization). The pre-trained MLP reproduces the characteristic separation between the thin and thick disks, as also demonstrated in \citet{2024ApJS..273...19Z}, whereas the DESI SP pipeline and models trained from scratch show no clear bimodality (see Appendix~\ref{sec:lr_scratch} for the effect of learning rate on the latter). One reason for our choice of sample selection is to retain as many sources as possible, so that the larger sample better highlights the contrast between the high- and low-[$\alpha$/Fe] sequences; the same [$\alpha$/Fe]–[Fe/H] comparisons using a clean, calibrated DESI SP subset are provided in Appendix~\ref{sec:compare_clean}.

We quantify the similarity between stellar chemical abundance distributions using the Maximum Mean Discrepancy (MMD; \citealt{JMLR:v13:gretton12a}), a non-parametric, kernel-based metric. The unbiased estimator of $\mathrm{MMD}^2$ is computed as
\begin{equation}
\begin{aligned}
\widehat{\mathrm{MMD}}^2 =\;&
\frac{1}{m(m-1)} \sum_{i \neq j}^{m} k(x_i, x_j)
+ \frac{1}{n(n-1)} \sum_{i \neq j}^{n} k(y_i, y_j) \\
&-
\frac{2}{mn} \sum_{i=1}^{m} \sum_{j=1}^{n} k(x_i, y_j)
\end{aligned}
\end{equation}
where $\{x_i\}_{i=1}^m$ and $\{y_i\}_{i=1}^n$ are samples drawn from two distributions in the two-dimensional $[\alpha/\mathrm{Fe}]$--$[\mathrm{Fe/H}]$ space. One distribution corresponds to one of the first four columns in Figure~\ref{fig:compare_pretrained_desi_afe_feh}, while the other corresponds to the APOGEE reference sample shown in the last column.

We adopt a Gaussian (RBF) kernel,
$k(x,y) = \exp(-\gamma \|x-y\|^2)$,
with bandwidth parameter $\gamma = 2.0$. Lower MMD values indicate greater similarity between the two distributions. As shown in Figure~\ref{fig:compare_pretrained_desi_afe_feh}, the pre-trained MLP models yield smaller MMD values than both the DESI SP pipeline and models trained from scratch, indicating improved consistency with the APOGEE abundance distribution.

Despite these improvements, Figure~\ref{fig:compare_pretrained_desi} shows that the pre-trained MLP systematically underestimates [Fe/H] in the metal-rich regime while slightly overestimating it in the metal-poor regime. For [$\alpha$/Fe], there is an overall tendency toward underestimation—particularly evident in Figure~\ref{fig:compare_pretrained_desi_afe_feh}, where the underestimation bias of [$\alpha$/Fe] is most pronounced in the metal-poor regime when compared with the DESI DR1–APOGEE DR17 cross-matched sample shown in the last column. These discrepancies likely reflect differences in resolution and signal-to-noise between LAMOST LRS and DESI spectra, and suggest that modest fine-tuning on DESI data may further improve performance.

\begin{table*}[htbp]
    \centering
    \caption{Coefficient of determination ($R^2$) for MLPs with different fine-tuning strategies and input types in [Fe/H] and [$\alpha$/Fe] estimation on DESI spectra.}
    \label{tab:finetune_embed_finetune}
    \begin{tabular}{l|cccc}
    \hline
    \multicolumn{5}{c}{\textbf{[Fe/H]}} \\
    \hline
     (Full/MP/MR) & lrs & embed & embed-align & embed-split \\
    \hline
    residual & 0.923 / \colorbox{blue!6}{0.636} / 0.904 & 0.922 / 0.196 / 0.913 & 0.922 / 0.003 / 0.918 & \colorbox{blue!24}{0.933} / 0.219 / \colorbox{blue!6}{0.927} \\
    LoRA & 0.922 / 0.571 / 0.904 & 0.921 / 0.031 / 0.916 & \colorbox{blue!12}{0.928} / 0.131 / 0.922 & \colorbox{blue!24}{0.933} / 0.133 / \colorbox{blue!12}{0.930} \\
    full fine-tune & 0.923 / 0.329 / 0.911 & 0.926 / 0.173 / 0.918 & \colorbox{blue!6}{0.927} / 0.085 / 0.922 & \colorbox{blue!24}{0.933} / 0.079 / \colorbox{blue!24}{0.931} \\
    \hline
    zero-shot & 0.873 / \colorbox{blue!12}{0.678} / 0.833 & 0.833 / \colorbox{blue!24}{0.694} / 0.779 & 0.851 / 0.465 / 0.810 & 0.839 / 0.537 / 0.791 \\
    from scratch & 0.880 / -0.736 / 0.878 & 0.877 / -1.246 / 0.886 & 0.647 / -0.072 / 0.542 & 0.794 / -0.649 / 0.759 \\
    \hline
    \multicolumn{5}{c}{\textbf{[$\alpha$/Fe]}} \\
    \hline
residual & 0.799 / 0.366 / 0.802 & 0.790 / \colorbox{blue!6}{0.396} / 0.792 & 0.796 / 0.232 / 0.801 & 0.796 / 0.335 / 0.800 \\
LoRA & \colorbox{blue!24}{0.829} / \colorbox{blue!24}{0.418} / \colorbox{blue!24}{0.831} & 0.807 / 0.370 / 0.810 & 0.769 / 0.290 / 0.772 & 0.733 / 0.205 / 0.736 \\
full-finetune & \colorbox{blue!12}{0.816} / 0.381 / \colorbox{blue!12}{0.819} & \colorbox{blue!6}{0.810} / 0.315 / \colorbox{blue!6}{0.813} & 0.784 / 0.247 / 0.787 & 0.773 / 0.231 / 0.777 \\
\hline
zero-shot & 0.776 / 0.318 / 0.778 & 0.730 / 0.183 / 0.733 & 0.772 / 0.202 / 0.776 & 0.681 / 0.105 / 0.685 \\
from scratch & 0.761 / 0.067 / 0.766 & 0.729 / \colorbox{blue!12}{0.401} / 0.730 & 0.611 / 0.144 / 0.612 & 0.749 / 0.374 / 0.751 \\
    \hline
    
    \end{tabular}
    \begin{flushleft}
        \textit{Note.} 
        \texttt{lrs} refers to an MLP trained directly on normalized LAMOST LRS. \texttt{embed} uses embeddings from the LAMOST LRS pre-trained encoder of SpecCLIP, \texttt{embed-align} uses projected embeddings of LAMOST LRS from the alignment between LAMOST LRS and Gaia XP spectra, and \texttt{embed-split} uses combined (shared + modality-specific) projected embeddings. \texttt{Residual}, \texttt{LoRA}, and \texttt{full fine-tune} denote different fine-tuning strategies applied to pre-trained MLPs. \texttt{zero-shot} indicates applying pre-trained MLPs directly to DESI spectra without fine-tuning, and \texttt{scratch} indicates MLPs trained from scratch on the same number of DESI spectra as used for fine-tuning. Reported values are $R^2$ formatted as full/metal-poor/metal-rich (Full/MP/MR), where metal-poor corresponds to [Fe/H] $<-1.0$, and metal-rich corresponds to [Fe/H] $>-1.0$. Top three entries of $R^2$ within each subset (``full'', ``poor'', and ``rich'') are highlighted with decreasing shades of blue. The numbers are reported as the average over 5 independent runs. Similar to Figure~\ref{fig:compare_pretrained_desi}, the number of test samples is 5,539, with a fine-tuning set of 2,069 stars. 
        \end{flushleft}
    \end{table*}
    
\subsection{Fine-tuning the MLP model}
\label{sec:fine_tune_mlp}

We next apply fine-tuning, using LoRA fine-tuning for [$\alpha$/Fe] and a residual-head strategy for [Fe/H] (See Section \ref{sec:fine_tune_strategies} for comparisons of fine-tuning strategies for the two parameters). The one-to-one comparisons in Figure~\ref{fig:compare_pretrained_desi} (\texttt{MLP-LRS (fine-tuned)}) show clear gains relative to the zero-shot models, largely correcting the anomalous trend noted above, which implies that fine-tuning on DESI helps compensate for differences in resolution and spectral properties that are not fully captured by pre-training on LAMOST LRS.

The [$\alpha$/Fe]–[Fe/H] diagram also improves after fine-tuning (Figure~\ref{fig:compare_pretrained_desi_afe_feh}, \texttt{MLP-LRS (fine-tuned)}). In particular, the underestimation of [$\alpha$/Fe] in the metal-poor regime is reduced, bringing the distribution into closer agreement with that of the DESI DR1–APOGEE DR17 cross-matched sample. The MMD value decreases further, indicating an additional improvement in the agreement between the model-predicted abundance distribution and the APOGEE reference sample.

Figure~\ref{fig:sample_effects} shows the effect of fine-tuning sample size on estimation performance after averaging over five independent runs\footnote{Independence is controlled by five random seeds that are fixed across experiments. These seeds determine the train/validation split and neural network configurations (e.g., weight initialization and dropout). For all sample sizes except 2,069, the seeds also determine which stars are selected for training and validation out of the total 2,069 available, which can lead to larger variation in these experiments.}. Notably, even with only 200–500 fine-tuning samples, the performance metrics remain robust and approach those obtained with the maximum sample size. However, for the residual scatter $\sigma$ of [Fe/H], larger fine-tuning sample sizes (1,000–2,000) are needed to achieve overall performance comparable to or better than the zero-shot case. Further insight into the scatter plots at different sample sizes is provided in Figures~\ref{fig:sample_size_one2one} and \ref{fig:sample_size_2d}.

These results demonstrate that small to moderate amounts of fine-tuning data can enhance cross-survey generalization, even for simple MLP models.

\section{Discussion}
\label{sec:discussion}

\subsection{Foundation models for LAMOST LRS}
\label{sec:speclip}

We first assess the role of foundation models in stellar parameter estimation. We compare three MLP variants: one trained directly on the normalized LAMOST LRS (\texttt{lrs}), one trained on embeddings from the LAMOST LRS pre-trained encoder (\texttt{embed}), one trained on projected embeddings from the alignment of LAMOST LRS and Gaia XP spectra via contrastive training (\texttt{embed-align}), and one trained on combined (shared + modality-specific) projected embeddings (\texttt{embed-split}). The latter three models rely on embeddings derived from SpecCLIP pre-trained in a self-supervised way (see Section~\ref{sec:pretrained_foundation}).

Table~\ref{tab:finetune_embed_finetune} (see, e.g., the \texttt{residual} row for [Fe/H] and \texttt{LoRA} row for [$\alpha$/Fe]) summarizes the results. For [Fe/H], the zero-shot performance of the embedding-based MLPs (\texttt{embed}, \texttt{embed-align}, and \texttt{embed-split}) is not superior to that of MLPs pre-trained on LAMOST LRS itself, except in the metal-poor regime, where \texttt{embed} performs better--although these comparisons may be limited by the small testing sample size (79). Fine-tuning the embedding-based MLPs improves performance in the metal-rich regime ([Fe/H] $>-1$), as exemplified by the best metal-rich results achieved by \texttt{embed-split}, but the fine-tuning also degrades performance in the metal-poor regime. For [$\alpha$/Fe], embedding-based MLPs generally show lower accuracy. These results suggest that, at least for [Fe/H] and [$\alpha$/Fe], embeddings from SpecCLIP pre-trained on LAMOST LRS and Gaia XP capture some information that downstream MLPs can readily exploit for cross-survey transfer to DESI. However, they may be less informative than the normalized LRS spectra themselves in certain parameter regimes when transferring to DESI, such as the metal-poor regime. A more detailed analysis of sample-size effects on different MLP input types (normalized spectra vs. embeddings) is provided in Appendix~\ref{sec:additional_tables}.

\subsection{Fine-tuning strategies}
\label{sec:fine_tune_strategies}

\begin{figure*}
    \begin{center}
    \includegraphics[scale=0.27,angle=0]{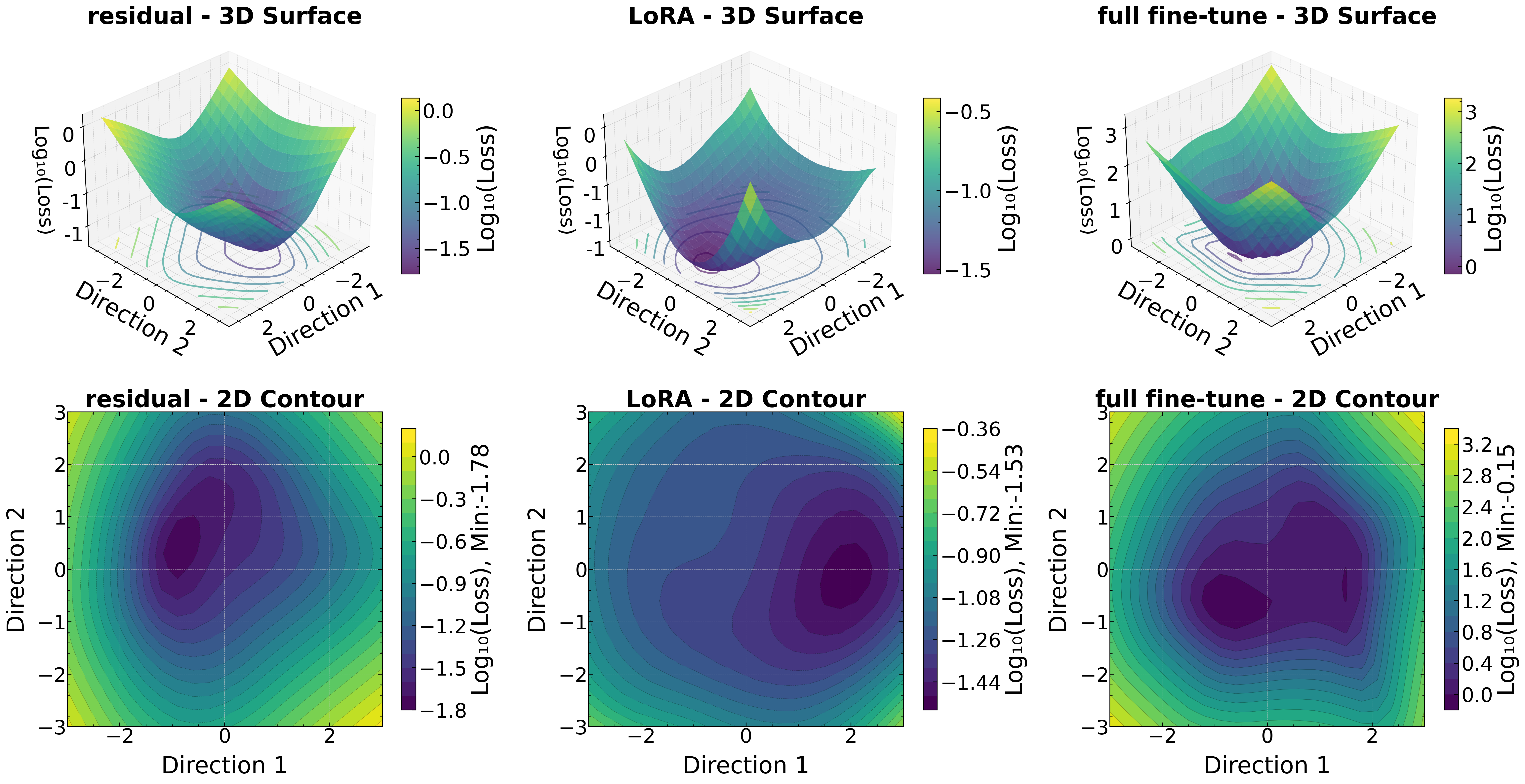} 
    \caption{Loss landscapes for different fine-tuning strategies, evaluated on the metal-poor regime of [Fe/H] (79 stars). The first row shows 3D surfaces, and the second row shows 2D contours of the logarithmic MSE loss, plotted as a function of perturbations around the pre-trained model parameters along two random directions. The minimum losses are indicated in the colorbar of the 2D contours.}
    \label{fig:loss_landscapes}
    \end{center}
\end{figure*}

Table~\ref{tab:finetune_embed_finetune} also compares different fine-tuning strategies on DESI spectra, including LoRA, residual-head, and full fine-tuning (see, e.g., the \texttt{lrs} column). Residual-head fine-tuning yields the best performance among the three strategies for [Fe/H] in the metal-poor regime, close to that of the zero-shot model. It shows slightly worse performance in the metal-rich regime compared to the other two strategies. Nevertheless, its metal-rich performance remains superior to the zero-shot results and corrects the underestimation trend, particularly around [Fe/H] $\sim0$, seen in Figure~\ref{fig:compare_pretrained_desi} (\texttt{MLP-LRS (fine-tuned)} vs. \texttt{MLP-LRS}). This suggests that residual heads are effective when spectral features are relatively clean and already well captured by the pre-trained MLP. This includes the metal-poor regime, since the pre-training set contained a relatively well-constructed metal-poor subset \citep{2025arXiv250701939Z}, so that only small output-level corrections are needed. In contrast, both LoRA and full fine-tuning tend to improve the metal-rich regime but degrade performance in the metal-poor regime, consistent with the LoRA behavior reported by \citet{2025arXiv250720972Z}. This likely reflects the fact that LoRA and full fine-tuning adjust internal representations, which may overfit to the dominant metal-rich ([Fe/H] $>-1.0$) population, at the expense of performance in the minority metal-poor regime.

For [$\alpha$/Fe], LoRA and full fine-tuning performs best. One plausible explanation is that [$\alpha$/Fe] is a composite quantity combining several elemental abundances, and its reliable estimation is more sensitive to differences in signal-to-noise and resolution between LAMOST LRS and DESI. Moreover, the distribution of [$\alpha$/Fe] is relatively balanced, which reduces the risk of overfitting to any particular regime. In such cases, more flexible adaptation, such as full fine-tuning, is beneficial. Since LoRA fine-tuning achieves slightly better performance across experiments, we adopt this strategy for [$\alpha$/Fe] in the main results of this paper. A more detailed analysis of sample-size effects on these strategies is provided in Appendix~\ref{sec:additional_tables}.

\subsection{Loss landscapes of fine-tuning strategies}
\label{sec:loss_landscapes}

To better understand the behaviors of different fine-tuning strategies, we examine their loss landscapes for [Fe/H] in the metal-poor regime, where fine-tuning is particularly challenging. Following \citet{2017arXiv171209913L}, we project the high-dimensional trainable parameter space--consisting of only the adapter weights for residual-head and LoRA, or all weights for full fine-tuning--onto a two-dimensional plane defined by two random directions $\delta$ and $\eta$ sampled from a random Gaussian distribution, and evaluate the loss function $f(\alpha,\beta) = L(\theta^* + \alpha \delta + \beta \eta)$, where $\theta^*$ is the set of pre-trained parameters and $\alpha$ and $\beta$ are scale coefficients along the two directions. We use a 21$\times$21 grid with $\alpha,\beta \in [-3,3]$ (corresponding to a maximum perturbation of three times the filter-wise L2 norm of weights), and normalize by filter-wise scaling, where a ``filter'' corresponds to the weights associated with one output neuron\footnote{For LoRA layers, the ``filter'' in our implementation corresponds to one rank vector in the LoRA-$A$ matrix, or one output-dimension vector in the LoRA-$B$ matrix.} in the MLP.

We adopt mean squared error (MSE) loss between predicted and true [Fe/H], computed on 79 DESI spectra (from the testing sample) with [Fe/H] $<-1$, labels taken from APOGEE DR17, using the \texttt{lrs} experiment. The resulting landscapes along two example random directions are shown in Figure~\ref{fig:loss_landscapes}. Compared with residual-head fine-tuning, LoRA yields a higher minimum in the landscape over the perturbations, and this minimum lies farther from the pre-trained solution, suggesting overfitting to the metal-rich regime, where the globally optimized weights are suboptimal for metal-poor stars. Full fine-tuning yields its minimum away from the pre-trained weights as well, but with an overall shallower loss landscape around the pre-trained solution. These trends are robust across five random seeds (ten random directions), with only one exception in which LoRA attains the smallest minimum--though this minimum occurs much farther from the pre-trained solution than in the residual-head fine-tuning case. These comparisons help explain the empirical differences observed across strategies. A more physically interpretable saliency analysis of spectral features, together with a comparison of different fine-tuning strategies in this regime, is provided in Appendix~\ref{sec:sensitivity}. In particular, the residual-head fine-tuning shows higher sensitivity of [Fe/H] to the relevant lines in the spectra. A detailed mathematical analysis of these discrepancies among fine-tuning strategies is left for future work.

\subsection{Limitations}
\label{sec:limitations}

Several limitations of this work should be noted. First, the spectral foundation models used here (SpecCLIP) are trained only on the blue region (400–560~nm) of LAMOST LRS. While this region contains many informative lines, it does not capture the full information content of stellar spectra. Second, the encoders we employ are relatively modest in scale ($\sim$42.7 million parameters for the LAMOST LRS model and $\sim$0.1 billion when including alignment with the Gaia XP model), whereas larger models may be required to represent more complex spectral features. Third, we demonstrate our analysis on only two parameters, [Fe/H] and [$\alpha$/Fe]; other quantities such as $T_{\rm eff}$ and $\log g$ may require different modeling choices. Fourth, the fine-tuning datasets are limited in size; larger samples may be necessary to fully evaluate the potential of foundation models, especially when fine-tuning backbone transformer layers rather than only downstream MLPs. In such cases, approaches like LoRA--shown to be effective in large language models and requiring only a limited number of trainable parameters--may prove more advantageous for spectral applications. Finally, while our results highlight the surprising strength of simple MLPs in cross-survey generalization, a foundation model trained directly on DESI spectra—with sufficient high-quality labels—may ultimately surpass them. Indeed, \citet{2025arXiv250701939Z} has shown that such models can achieve state-of-the-art performance, suggesting that future survey-specific foundation models remain a promising direction.

\section{Summary}
\label{sec:summary}

We summarize the main findings of this work as follows. This paper investigates the effectiveness of pre-trained multilayer perceptron (MLP) models for low-to-moderate-resolution cross-survey generalization, using the transfer from LAMOST low-resolution spectra (LRS) to DESI medium-resolution spectra (MRS) as a case study. 

First, we show that pre-trained MLPs produce estimates that are more consistent with APOGEE labels than the DESI stellar parameter pipeline for both iron abundance ([Fe/H]) and $\alpha$-elemental abundance ([$\alpha$/Fe]), even in a zero-shot setting without fine-tuning. Notably, the model recovers the two thin-thick disk sequences in the DESI DR1 sample--features that do not appear in \citet{2025arXiv250514787K}. Modest fine-tuning ($\sim$2,000 spectra) with DESI further improves performance, suggesting that fine-tuning helps mitigate differences in resolution and data quality between surveys. 

Second, we examine the role of foundation models. While embeddings from transformer-based spectral foundation models improve [Fe/H] performance in the metal-rich regime, they degrade performance in the metal-poor regime and do not improve [$\alpha$/Fe] predictions. This indicates that, for cross-survey transfer between LAMOST and DESI, simple MLPs trained directly on LRS remain competitive with more complex foundation models. 

Third, we find that fine-tuning strategies have parameter-dependent effects: residual-head fine-tuning is most effective for [Fe/H], while LoRA yields better results for [$\alpha$/Fe] given our fine-tuning sample size. These findings highlight the importance of tailoring fine-tuning approaches to the specific stellar parameter of interest. 

Overall, our results demonstrate that pre-trained MLPs offer a strong and practical baseline for cross-survey generalization in stellar spectroscopy. These models highlight how even lightweight architectures can capture robust information transfer across surveys, while the role of larger spectral foundation models remains an open direction for exploration--especially as broader wavelength coverage, increased model capacity, and larger fine-tuning datasets become available.

From a survey design perspective, these findings suggest a practical path for upcoming large-scale stellar spectroscopic programs such as the stellar program with DESI, SDSS-V, and the Galactic Archaeology survey of the Subaru Strategic Program with the Prime Focus Spectrograph on the Subaru telescope \citep{PFS2026}. Simple pre-trained MLPs, when initialized on existing low-resolution surveys and fine-tuned with modest samples of new survey data, can already deliver competitive parameter estimates and recover key Galactic structures such as the thin–thick disk separation. This demonstrates the immediate utility of lightweight models for rapid deployment.

\section*{Acknowledgements}
X.Z., R.F.G.W., A.S.S., L.D., T.B., and V.W. have been supported by a grant from the Schmidt Sciences. Y.S.T is supported by the National Science Foundation under Grant No. AST-2406729. 
Y.H. acknowledges the support from the National Science Foundation of China (NSFC grant No. 12422303), the Fundamental Research Funds for the Central Universities (grant Nos. 118900M122, E5EQ3301X2, and E4EQ3301X2), and the National Key R\&D Programme of China (grant No. 2023YFA1608303).

This work made use of the data from LAMOST (Large Sky Area Multi-Object Fiber Spectroscopic Telescope, also known as the Guoshoujing Telescope) (https://cstr.cn/31118.02.LAMOST). LAMOST is a Chinese national mega-science facility, operated by National Astronomical Observatories, Chinese Academy of Sciences.

This research used data obtained with the Dark Energy Spectroscopic Instrument (DESI). DESI construction and operations is managed by the Lawrence Berkeley National Laboratory. This material is based upon work supported by the U.S. Department of Energy, Office of Science, Office of High-Energy Physics, under Contract No. DE–AC02–05CH11231, and by the National Energy Research Scientific Computing Center, a DOE Office of Science User Facility under the same contract. Additional support for DESI was provided by the U.S. National Science Foundation (NSF), Division of Astronomical Sciences under Contract No. AST-0950945 to the NSF’s National Optical-Infrared Astronomy Research Laboratory; the Science and Technology Facilities Council of the United Kingdom; the Gordon and Betty Moore Foundation; the Heising-Simons Foundation; the French Alternative Energies and Atomic Energy Commission (CEA); the National Council of Humanities, Science and Technology of Mexico (CONAHCYT); the Ministry of Science and Innovation of Spain (MICINN), and by the DESI Member Institutions: www.desi.lbl.gov/collaborating-institutions. The DESI collaboration is honored to be permitted to conduct scientific research on I’oligam Du’ag (Kitt Peak), a mountain with particular significance to the Tohono O’odham Nation. Any opinions, findings, and conclusions or recommendations expressed in this material are those of the author(s) and do not necessarily reflect the views of the U.S. National Science Foundation, the U.S. Department of Energy, or any of the listed funding agencies.

Funding for the Sloan Digital Sky Survey V has been provided by the Alfred P. Sloan Foundation, the Heising-Simons Foundation, the National Science Foundation, and the Participating Institutions. SDSS acknowledges support and resources from the Center for High-Performance Computing at the University of Utah. SDSS telescopes are located at Apache Point Observatory, funded by the Astrophysical Research Consortium and operated by New Mexico State University, and at Las Campanas Observatory, operated by the Carnegie Institution for Science. The SDSS web site is www.sdss.org. 

SDSS is managed by the Astrophysical Research Consortium for the Participating Institutions of the SDSS Collaboration, including the Carnegie Institution for Science, Chilean National Time Allocation Committee (CNTAC) ratified researchers, Caltech, the Gotham Participation Group, Harvard University, Heidelberg University, The Flatiron Institute,  The Johns Hopkins University, L’Ecole polytechnique fédérale de Lausanne (EPFL), Leibniz-Institut für Astrophysik Potsdam (AIP), Max-Planck-Institut für Astronomie (MPIA Heidelberg), Max-Planck-Institut für Extraterrestrische Physik (MPE), Nanjing University, National Astronomical Observatories of China (NAOC), New Mexico State University, The Ohio State University, Pennsylvania State University, Smithsonian Astrophysical Observatory, Space Telescope Science Institute (STScI), the Stellar Astrophysics Participation Group, Universidad Nacional Autónoma de México (UNAM), University of Arizona, University of Colorado Boulder, University of Illinois at Urbana-Champaign, University of Toronto, University of Utah, University of Virginia, Yale University, and Yunnan University.

\section*{Data Availability}
All observational data used in this work are publicly available from the archives. No proprietary data were used. The code and trained weights are available upon request.

\software{
    SpecCLIP \citep{zhao_2025_17824840},
    PyTorch \citep{NEURIPS2019_9015},
    Astropy \citep{2013A&A...558A..33A, 2018AJ....156..123A},
    SPARCL \citep{2024arXiv240105576J}
}

\appendix
\restartappendixnumbering

\section{Ablation Studies}
\label{sec:ablation}

In this section, we systematically examine the impact of key fine-tuning factors, including the number of training samples, the number of trainable parameters, and the learning rate for training from scratch. The results are presented in Appendices A.1–A.3.

\subsection{Scatter plots and performance of embedding types and fine-tuning strategies across sample sizes}
\label{sec:additional_tables}

Here we present additional results on how fine-tuning sample size impacts one-to-one comparisons with APOGEE labels, and on the performance of different embedding types and fine-tuning strategies.

\begin{figure*}[h]
        \begin{center}
        \includegraphics[scale=0.175,angle=0]{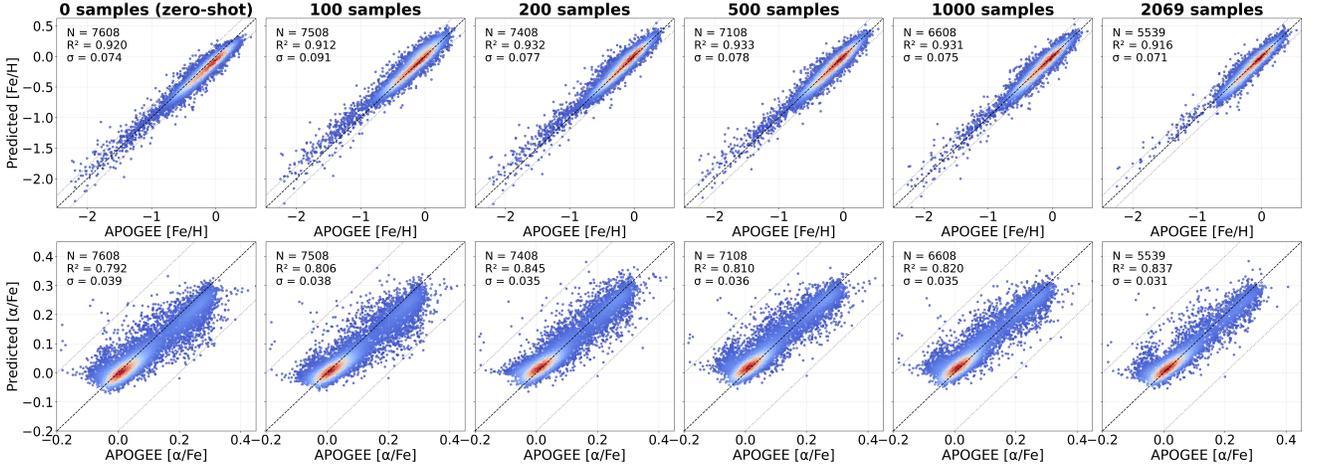} 
        \caption{Effect of fine-tuning sample size on the one-to-one comparisons, evaluated on all remaining samples not used for fine-tuning (See Table \ref{tab:embed_ratio} for comparisons using a fixed testing sample). Results are shown for residual-head fine-tuning of [Fe/H] and LoRA fine-tuning of [$\alpha$/Fe]. Other conventions follow Figure~\ref{fig:compare_pretrained_desi}.}
        \label{fig:sample_size_one2one}
        \end{center}
\end{figure*}

\begin{figure*}[h]
        \begin{center}
        \includegraphics[scale=0.14,angle=0]{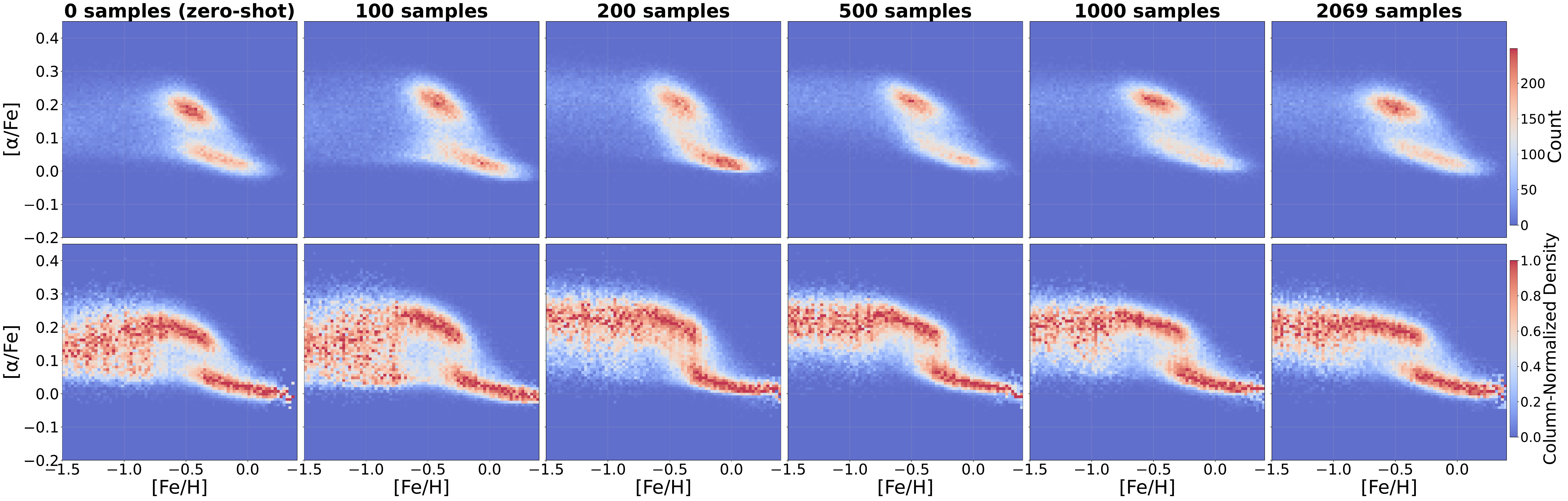}
        \caption{Effect of fine-tuning sample size on the predicted [$\alpha$/Fe]–[Fe/H] diagrams, evaluated on all remaining samples not used for fine-tuning. Results are shown for residual-head fine-tuning of [Fe/H] and LoRA fine-tuning of [$\alpha$/Fe]. Other conventions follow Figure~\ref{fig:compare_pretrained_desi_afe_feh}.}
        \label{fig:sample_size_2d}
        \end{center}
\end{figure*}

Figure~\ref{fig:sample_size_one2one} and \ref{fig:sample_size_2d} show the one-to-one comparisons between the predictions and the APOGEE labels and the [$\alpha$/Fe]–[Fe/H] diagrams across different fine-tuning sample sizes. The evaluation is performed on all remaining samples not used for fine-tuning (in contrast to all other tables and figures in this paper, which use only the 5,539-star testing sample), in order to maximize the testing coverage in each case. For [Fe/H], the underestimation trend in the metal-rich regime is already corrected with as few as 100 fine-tuning samples, although the overall scatter is initially larger and decreases as the fine-tuning sample size increases. For [$\alpha$/Fe], the underestimation trend begins to be corrected with about 200 fine-tuning samples and continues to improve with larger sample sizes. Note that $R^2$ fluctuates because it is affected by both the testing sample size and random-seed variance. Comparisons using the fixed 5,539-star testing sample, with results averaged over the same five seeds, are shown in Table \ref{tab:embed_ratio}.

Table~\ref{tab:embed_ratio} compares MLPs trained on normalized spectra (\texttt{lrs}), SpecCLIP embeddings (\texttt{embed}, \texttt{embed-align}, and \texttt{embed-split}), and models trained from scratch across a range of fine-tuning sample sizes, for both [Fe/H] and [$\alpha$/Fe]. The ratio of metal-poor to metal-rich stars is kept fixed in each subset. Using the \texttt{lrs} model with as few as 100 (12:88) fine-tuning samples, the pre-trained MLP achieves better performance in the majority of cases (metal-rich stars for [Fe/H] and $\alpha$-rich stars for [$\alpha$/Fe]) compared to the zero-shot case (last column). This demonstrates that the pre-trained MLP has learned transferable features from the LAMOST LRS dataset that can be effectively adapted to DESI with only small amounts of fine-tuning data. Embedding-based MLPs perform better in the metal-rich regime across sample sizes, but underperform in the metal-poor regime for [Fe/H] and consistently underperform for [$\alpha$/Fe]. Models trained from scratch perform poorly at all sample sizes, underscoring the importance of pre-training.

Table~\ref{tab:finetune_ratio} evaluates fine-tuning strategies under the same sample-size variations. For [Fe/H], residual-head fine-tuning consistently yields the best results among the three strategies, particularly in the metal-poor regime, and remains robust with small sample sizes. It also shows better performance in the metal-rich regime when the fine-tuning sample size is limited. LoRA and full fine-tuning perform competitively in the metal-rich regime with moderate fine-tuning sample sizes, but perform less well with smaller sample sizes and in the metal-poor regime. For [$\alpha$/Fe], LoRA fine-tuning performs best overall, with full fine-tuning remaining competitive and achieving comparable performance, while residual-head fine-tuning performs well in the metal-rich regime for several smaller sample sizes. In all cases, models trained from scratch perform poorly, again highlighting the benefits of pre-training. 

\begin{table}[H]
    \centering
    \caption{Coefficient of determination ($R^2$) for MLPs with different input types as a function of fine-tuning sample size.}
    \label{tab:embed_ratio}
    \begin{tabular}{l|ccc}
        \hline
        \hline
        \multicolumn{4}{c}{\textbf{[Fe/H]}} \\
        \hline
    M-poor:M-rich & 243:1826 & 117:883 & 59:441 \\
    \hline
    lrs & 0.923 / 0.636 / 0.904 & 0.916 / 0.619 / 0.894 & 0.910 / 0.600 / 0.886 \\
    embed & 0.922 / 0.196 / 0.913 & 0.920 / 0.160 / 0.911 & 0.913 / 0.216 / 0.900 \\
    embed-align & 0.922 / 0.003 / \colorbox{blue!6}{0.918} & 0.922 / -0.026 / \colorbox{blue!6}{0.918} & 0.910 / -0.233 / 0.907 \\
    embed-split & \colorbox{blue!24}{0.933} / 0.219 / \colorbox{blue!24}{0.927} & \colorbox{blue!12}{0.927} / 0.058 / \colorbox{blue!12}{0.924} & \colorbox{blue!6}{0.926} / 0.213 / 0.917 \\
    \hline
    from scratch & 0.880 / -0.736 / 0.878 & 0.826 / -1.663 / 0.826 & 0.433 / -3.575 / 0.336 \\
    \hline
    M-poor:M-rich & 23:177 & 12:88 & 0:0 \\
    \hline
    lrs & 0.909 / 0.616 / 0.885 & 0.906 / \colorbox{blue!6}{0.643} / 0.881 & 0.873 / \colorbox{blue!12}{0.678} / 0.833 \\
    embed & 0.916 / 0.232 / 0.904 & 0.915 / 0.245 / 0.902 & 0.833 / \colorbox{blue!24}{0.694} / 0.779 \\
    embed-align & 0.914 / -0.247 / 0.913 & 0.914 / -0.150 / 0.911 & 0.851 / 0.465 / 0.810 \\
    embed-split & 0.925 / -0.098 / \colorbox{blue!12}{0.924} & 0.916 / -0.130 / 0.913 & 0.839 / 0.537 / 0.791 \\
    \hline
    from scratch & 0.098 / -4.912 / -0.090 & -0.575 / -8.525 / -0.924  & \\
    \hline
    \hline
    \multicolumn{4}{c}{\textbf{[$\alpha$/Fe]}} \\
    \hline
    M-poor:M-rich & 243:1826 & 117:883 & 59:441 \\
    \hline
    lrs & \colorbox{blue!24}{0.829} / \colorbox{blue!24}{0.418} / \colorbox{blue!24}{0.831} & \colorbox{blue!12}{0.828} / \colorbox{blue!6}{0.363} / \colorbox{blue!24}{0.831} & 0.820 / 0.287 / \colorbox{blue!12}{0.824} \\
embed & 0.807 / \colorbox{blue!12}{0.370} / 0.810 & 0.797 / 0.184 / 0.802 & 0.786 / 0.266 / 0.790 \\
embed-align & 0.769 / 0.290 / 0.772 & 0.762 / 0.212 / 0.765 & 0.779 / 0.194 / 0.784 \\
embed-split & 0.733 / 0.205 / 0.736 & 0.758 / 0.084 / 0.764 & 0.714 / 0.028 / 0.719 \\
    \hline
    from scratch & 0.761 / 0.067 / 0.766 & 0.647 / -0.042 / 0.651 & -0.333 / -0.880 / -0.342 \\
    \hline
    M-poor:M-rich & 23:177 & 12:88 & 0:0 \\
    \hline
    lrs & \colorbox{blue!6}{0.827} / 0.329 / \colorbox{blue!24}{0.831} & 0.808 / 0.297 / \colorbox{blue!6}{0.812} & 0.776 / 0.318 / 0.778 \\
embed & 0.793 / 0.298 / 0.796 & 0.785 / 0.265 / 0.789 & 0.730 / 0.183 / 0.733 \\
embed-align & 0.777 / 0.221 / 0.781 & 0.774 / 0.244 / 0.778 & 0.772 / 0.202 / 0.776 \\
embed-split & 0.725 / 0.089 / 0.730 & 0.725 / 0.121 / 0.729 & 0.681 / 0.105 / 0.685 \\
    \hline
    from scratch & -0.381 / -0.867 / -0.392 & -1.631 / -2.117 / -1.657 &  \\
    \hline
    \hline
    \end{tabular}
    \begin{flushleft}
        \small
        \textit{Note.} 
        Conventions follow Table~\ref{tab:finetune_embed_finetune}, but with varying numbers of fine-tuning samples (keeping the same ratio of metal-poor (M-poor) to metal-rich (M-rich) stars, as listed in the first row for [Fe/H] and [$\alpha$/Fe]). The last column shows the zero-shot results. Top three entries of $R^2$ within each subset (``full'', ``poor'', and ``rich'') are highlighted with decreasing shades of blue.
        \end{flushleft}
    \end{table}
    
\begin{table}[H]
\centering
\caption{Coefficient of determination ($R^2$) for MLPs with different fine-tuning strategies as a function of fine-tuning sample size.}
\label{tab:finetune_ratio}
\begin{tabular}{l|ccc}
\hline
\hline
\multicolumn{4}{c}{\textbf{[Fe/H]}} \\
\hline
M-poor:M-rich & 243:1826 & 117:883 & 59:441 \\
\hline
residual & \colorbox{blue!24}{0.923} / \colorbox{blue!6}{0.636} / \colorbox{blue!12}{0.904} & \colorbox{blue!6}{0.916} / 0.619 / 0.894 & 0.910 / 0.600 / 0.886 \\
LoRA & \colorbox{blue!12}{0.922} / 0.571 / \colorbox{blue!12}{0.904} & 0.912 / 0.496 / 0.891 & 0.914 / 0.590 / 0.892 \\
full fine-tune & \colorbox{blue!24}{0.923} / 0.329 / \colorbox{blue!24}{0.911} & \colorbox{blue!6}{0.916} / 0.401 / \colorbox{blue!6}{0.900} & 0.915 / 0.516 / 0.895 \\
\hline
scratch & 0.880 / -0.736 / 0.878 & 0.826 / -1.663 / 0.826 & 0.433 / -3.575 / 0.336 \\
\hline
M-poor:M-rich & 23:177 & 12:88 & 0:0 \\
\hline
residual & 0.909 / 0.616 / 0.885 & 0.906 / \colorbox{blue!12}{0.643} / 0.881 & 0.873 / \colorbox{blue!24}{0.678} / 0.833 \\
LoRA & 0.897 / 0.562 / 0.869 & 0.874 / 0.499 / 0.840 & 0.873 / \colorbox{blue!24}{0.678} / 0.833 \\
full fine-tune & 0.904 / 0.494 / 0.881 & 0.887 / 0.501 / 0.858 & 0.873 / \colorbox{blue!24}{0.678} / 0.833 \\
\hline
from scratch & 0.098 / -4.912 / -0.090 & -0.575 / -8.525 / -0.924 &  \\
\hline
\hline
\multicolumn{4}{c}{\textbf{[$\alpha$/Fe]}} \\
\hline
M-poor:M-rich & 243:1826 & 117:883 & 59:441 \\
\hline
residual & 0.799 / \colorbox{blue!6}{0.366} / 0.802 & 0.824 / 0.285 / 0.828 & \colorbox{blue!24}{0.840} / 0.347 / \colorbox{blue!24}{0.844} \\
LoRA & \colorbox{blue!6}{0.829} / \colorbox{blue!24}{0.418} / \colorbox{blue!6}{0.831} & 0.828 / 0.363 / \colorbox{blue!6}{0.831} & 0.820 / 0.287 / 0.824 \\
full-finetune & 0.816 / \colorbox{blue!12}{0.381} / 0.819 & 0.816 / 0.356 / 0.819 & 0.812 / 0.264 / 0.816 \\
\hline
from scratch & 0.761 / 0.067 / 0.766 & 0.647 / -0.042 / 0.651 & -0.333 / -0.880 / -0.342 \\
\hline
M-poor:M-rich & 23:177 & 12:88 & 0:0 \\
\hline
residual & 0.813 / 0.292 / 0.817 & \colorbox{blue!12}{0.830} / 0.206 / \colorbox{blue!12}{0.835} & 0.776 / 0.318 / 0.778 \\
LoRA & 0.827 / 0.329 / \colorbox{blue!6}{0.831} & 0.808 / 0.297 / 0.812 & 0.776 / 0.318 / 0.778 \\
full-finetune & 0.826 / 0.329 / 0.830 & 0.807 / 0.290 / 0.810 & 0.776 / 0.318 / 0.778 \\
\hline
scratch & -0.381 / -0.867 / -0.392 & -1.631 / -2.117 / -1.657 &  \\
\hline
\hline
\end{tabular}
\begin{flushleft}
    \small
    \textit{Note.} 
    Conventions follow Table~\ref{tab:finetune_embed_finetune}, but with varying numbers of fine-tuning samples (keeping the same ratio of metal-poor (M-poor) to metal-rich (M-rich) stars, as listed in the first row for [Fe/H] and [$\alpha$/Fe]). The last column shows the zero-shot results. Top three entries of $R^2$ within each subset (``full'', ``poor'', and ``rich'') are highlighted with decreasing shades of blue.
    \end{flushleft}
\end{table}

\subsection{Number of trainable parameters for different fine-tuning strategies}
\label{sec:trainable_params}

\begin{figure*}
        \begin{center}
        \includegraphics[scale=0.42,angle=0]{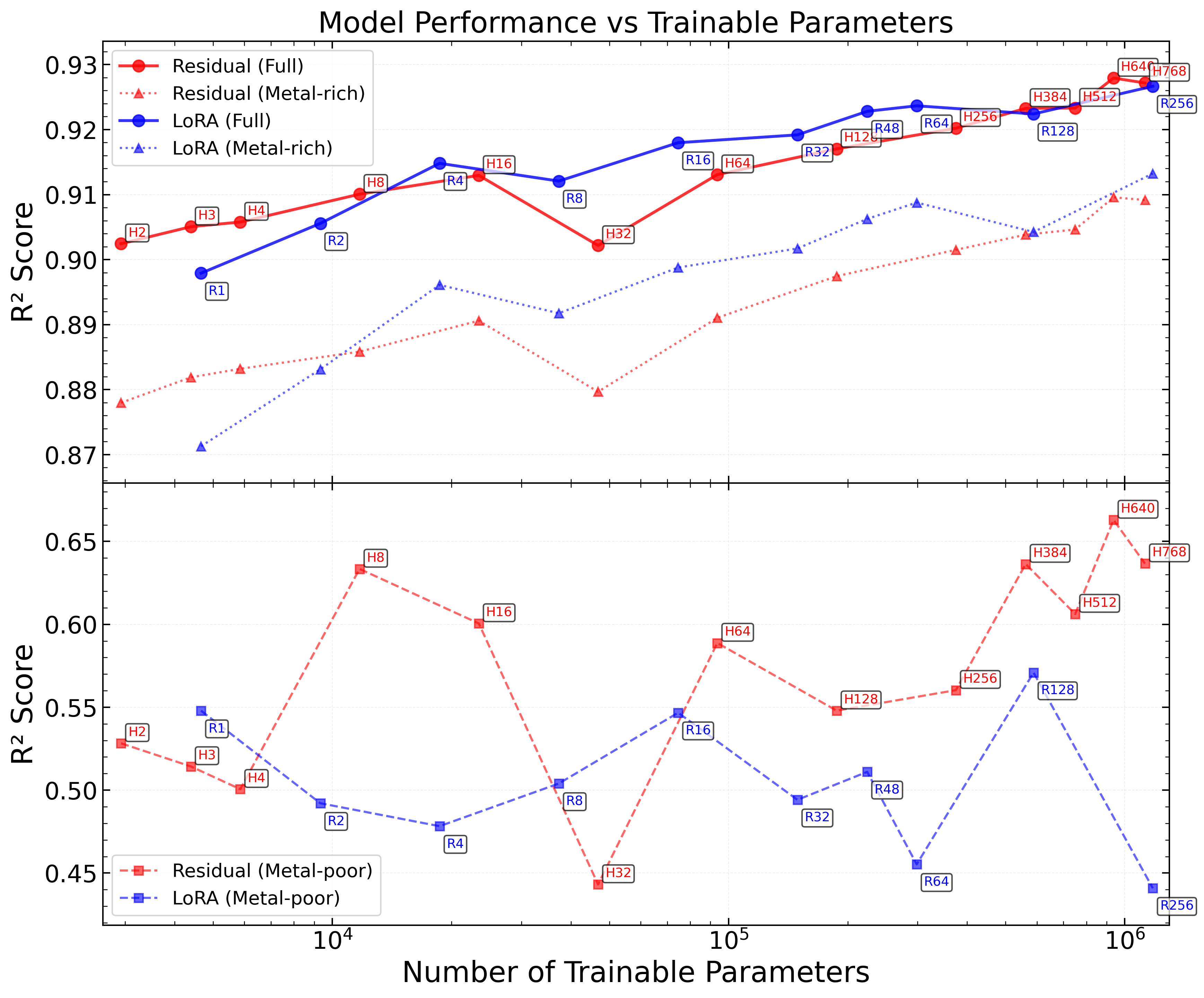} 
        \caption{Effect of the number of trainable parameters on [Fe/H] fine-tuning performance, comparing LoRA and residual-head strategies. The y-axis shows the coefficient of determination ($R^2$) for [Fe/H] estimation using the full sample, metal-rich subset ([Fe/H] $>-1.0$), and metal-poor subset ([Fe/H] $<-1.0$). The number of test samples is 5,539, with a fine-tuning set of 2,069 stars. Annotations indicate the hidden-layer size for residual-head models (e.g., \texttt{H384}) and the rank for LoRA models (e.g., \texttt{R128}).}
        \label{fig:trainable_params}
        \end{center}
\end{figure*}

We examine how the number of trainable parameters affects the performance of two parameter-efficient fine-tuning strategies: residual-head and LoRA. Figure~\ref{fig:trainable_params} shows the results for [Fe/H] estimation using the full set of 2,069 fine-tuning samples, averaged over five independent runs. LoRA achieves better performance in the metal-rich regime, whereas residual-head fine-tuning performs better in the metal-poor regime, across a range from a few thousand up to $\sim$1M trainable parameters. For the results reported in the main text, we adopt a hidden layer size of 384 (\texttt{H384}) for residual-head fine-tuning and a rank of 128 (\texttt{R128}) for LoRA, corresponding to $\sim$0.6 million trainable parameters for both strategies, ensuring a fair comparison.

\subsection{Learning rate for training from scratch}
\label{sec:lr_scratch}

\begin{figure*}
        \begin{center}
        \includegraphics[scale=0.23,angle=0]{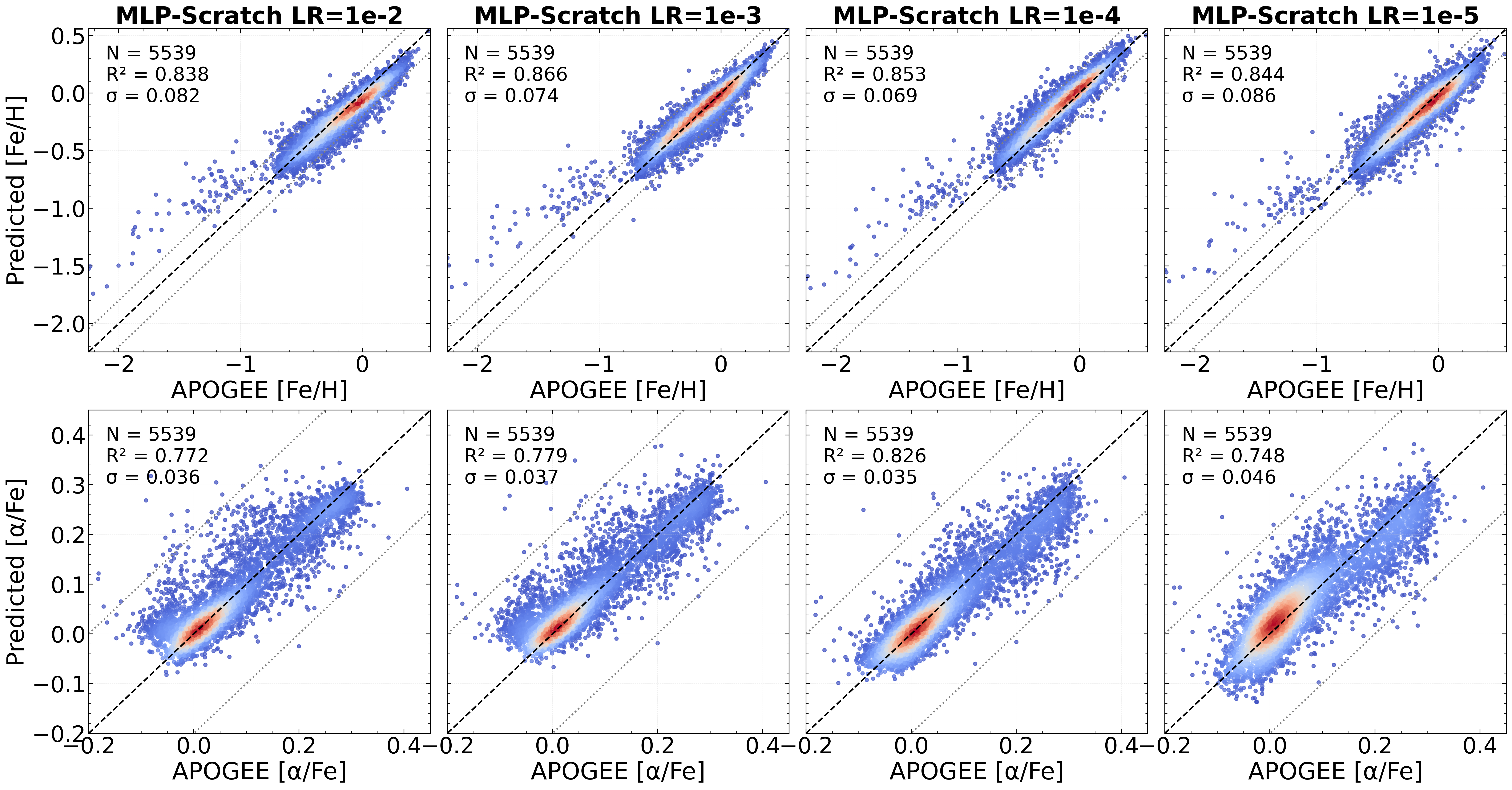} 
        \caption{Effect of learning rate (LR) on models trained from scratch for [Fe/H] and [$\alpha$/Fe], referenced against APOGEE. Conventions follow Figure~\ref{fig:compare_pretrained_desi}.}
        \label{fig:compare_pretrained_desi_lr}
        \end{center}
\end{figure*}

\begin{figure*}
        \begin{center}
        \includegraphics[scale=0.22,angle=0] {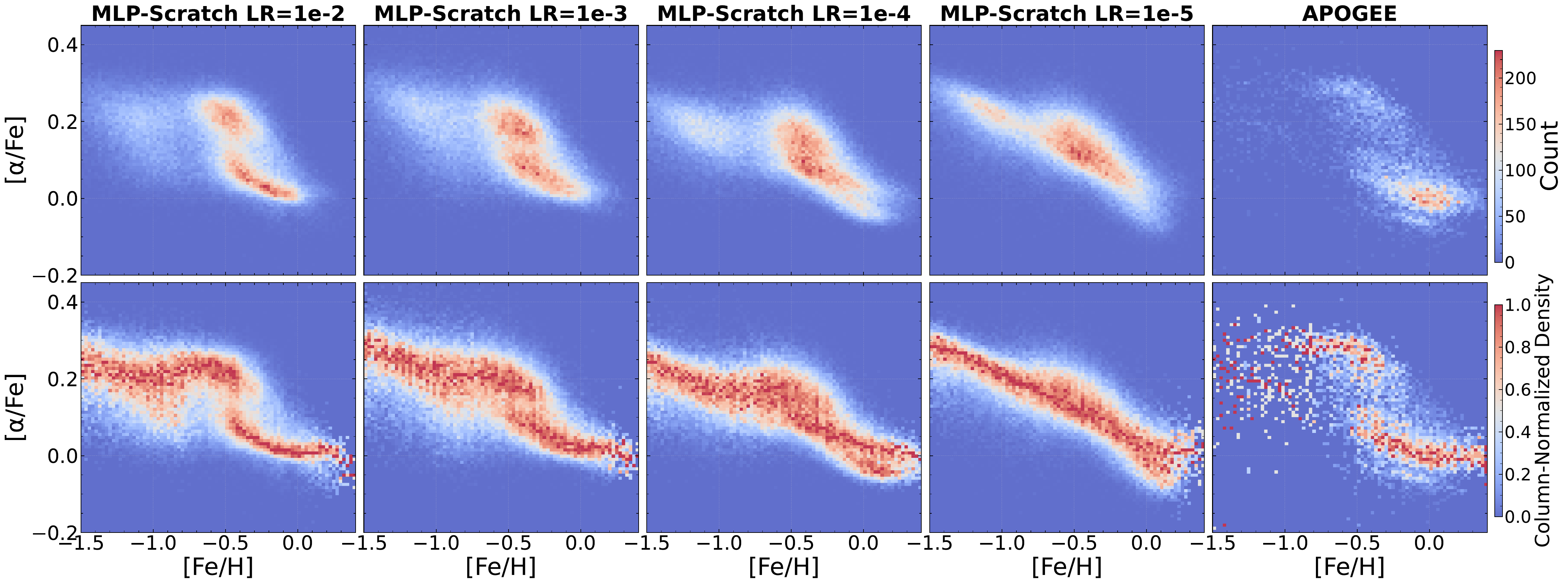}
        \caption{Effect of learning rate (LR) on models trained from scratch for the [$\alpha$/Fe]–[Fe/H] diagram. Conventions follow Figure~\ref{fig:compare_pretrained_desi_afe_feh}.}
        \label{fig:compare_pretrained_desi_afe_feh_lr}
        \end{center}
\end{figure*}

We also test the effect of learning rate on models trained from scratch with DESI spectra. Four different learning rates are explored (1e-2, 1e-3, 1e-4, and 1e-5), using the same number of DESI spectra as in the fine-tuning experiments. As shown in Figure~\ref{fig:compare_pretrained_desi_lr}, a relatively large learning rate (1e-3) yields better performance for both [Fe/H] and [$\alpha$/Fe] compared to the smaller rate (1e-5) used for LAMOST LRS pre-training and DESI fine-tuning. The [$\alpha$/Fe]–[Fe/H] diagram in Figure~\ref{fig:compare_pretrained_desi_afe_feh_lr} also indicates that a learning rate of 1e-3 better recovers the thin–thick disk separation, whereas 1e-5 fails to capture these trends. However, even this selected learning rate introduces systematic effects: [Fe/H] is overestimated in the metal-poor regime (Figure~\ref{fig:compare_pretrained_desi_lr}), and [$\alpha$/Fe] is likely overestimated for metal-poor stars (Figure~\ref{fig:compare_pretrained_desi_afe_feh_lr}).

\section{Saliency Analysis of Spectral Features}
\label{sec:sensitivity}

\begin{figure*}
        \begin{center}
        \includegraphics[scale=0.28,angle=0]{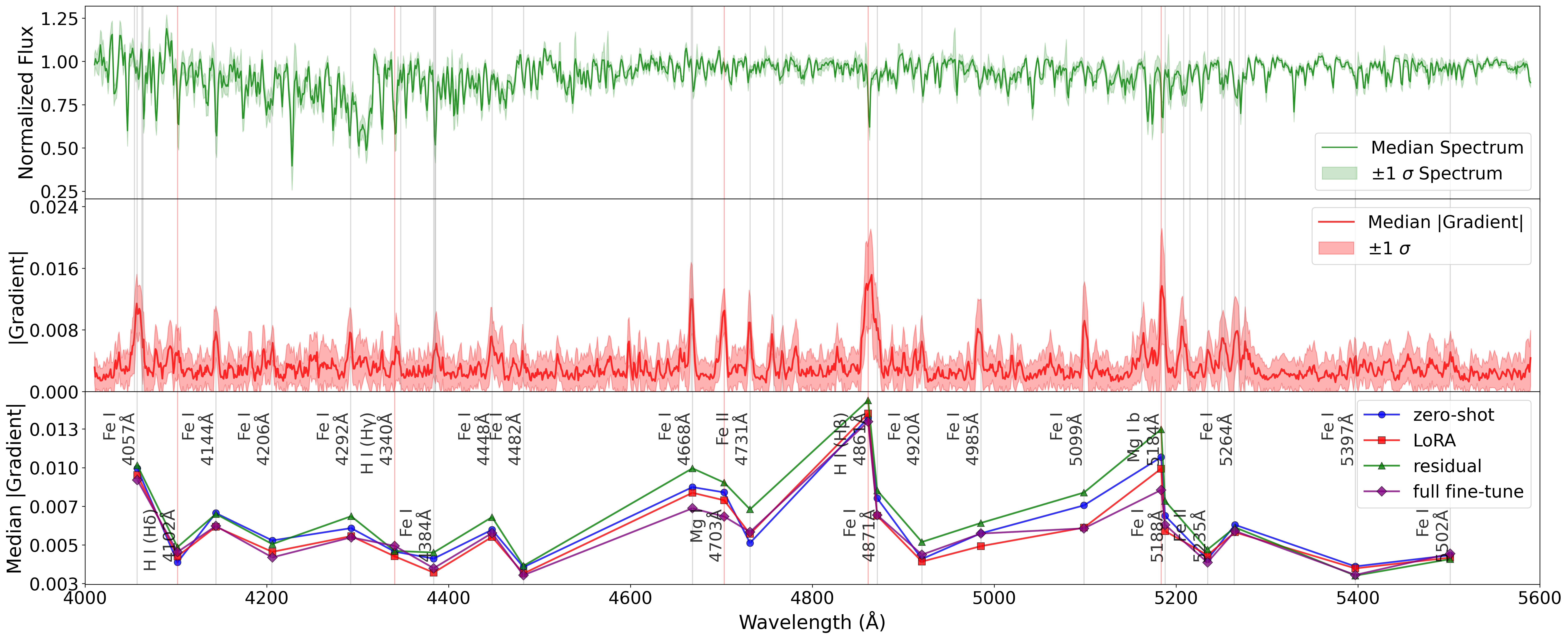} 
        \caption{Saliency maps of the output iron abundance with respect to the input spectra in the rest frame. The upper panel shows the median and 1$\sigma$ dispersion of the normalized spectra for the testing metal-poor subset ([Fe/H]). The middle panel presents the saliency maps (absolute gradients; median and 1$\sigma$) for the residual-head fine-tuning, indicating the sensitivity of the output iron abundance to different spectral regions. The lower panel highlights selected spectral lines with prominent sensitivity, comparing all fine-tuning strategies and the zero-shot results. Lines (from \citealp{kramida2024nist}) corresponding to major spikes are shown in grey (iron) and red (non-iron) in the upper and middle panels, and selectively plotted in the lower panel.}
        \label{fig:saliency}
        \end{center}
\end{figure*}

Here we examine the sensitivity of the output iron abundance--in particular for the metal-poor subset--to the input DESI spectra. The saliency maps \citep{Simonyan2014DeepIC} used here are defined as the gradients $\partial \widehat{[\mathrm{Fe/H}]}/\partial F_{\lambda}$, i.e.\ the derivative of the predicted abundance $\widehat{[\mathrm{Fe/H}]}$ with respect to the input spectrum $F_{\lambda}$. We take the absolute value of the gradients as a first step in visualization. The results are shown in Figure~\ref{fig:saliency}, where all spectra and saliency maps have been shifted to the rest frame using radial velocities reported by the \texttt{DESI SP} pipeline.  

Comparing the upper and middle panels, we find that the neural networks after residual-head fine-tuning are sensitive to both relatively clean and relatively blended iron lines. The lower panel shows that zero-shot testing with the MLP trained on LAMOST LRS already captures the correct sensitivity to iron lines in DESI spectra, even without ever having seen DESI data during training. Residual-head fine-tuning further enhances the sensitivity to most iron lines, which may partly explain its superior performance in [Fe/H] estimation (Sections~\ref{sec:fine_tune_strategies} and~\ref{sec:loss_landscapes}).  

An additional finding from the saliency maps is that the networks show the strongest sensitivity to the H$\beta$ Balmer line, with somewhat weaker but still notable sensitivity to H$\delta$ and H$\gamma$. The next strongest signal is to the Mg~I~b triplet, especially the 5184~\AA\ line, with smaller but non-negligible sensitivity to Mg~I~4703~\AA. After H$\beta$ and Mg~I~b, the iron lines appear as the next most sensitive features.  

This ranking implies that in the metal-poor regime--where iron lines are intrinsically weak--the neural networks first learn to constrain effective temperature ($T_{\rm eff}$) and surface gravity (log~$g$) from the Balmer lines, then infer overall metallicity from the Mg lines (which correlate with [Fe/H]), and finally refine the [Fe/H] estimate using the weaker iron lines themselves. This process mirrors traditional spectroscopic analysis by human experts, particularly in the metal-poor regime.  

These results demonstrate that the neural networks are not simply “black boxes” but can learn physically meaningful strategies for label inference such as [Fe/H]. They also suggest that joint training on additional labels (e.g., $T_{\rm eff}$, log~$g$, and [Mg/Fe]) could reduce degeneracies in the [Fe/H] gradients, so that the saliency maps would primarily highlight true iron lines, leading to more interpretable and reliable abundance estimations. A more detailed discussion of these topics, including the sign of the gradients, is left to future work.

\section{Dataset Selection Workflow}
\label{sec:dataset_cut}

A summary of the cuts for DESI DR1 and DESI EDR samples and resulting sample sizes, adopted in the main analysis, is provided in Table~\ref{tab:dataset_summary} (see also Section~\ref{sec:datasets}).

\begin{deluxetable}{lccc}
\tablecaption{Summary of DESI DR1 and DESI EDR samples adopted in the main analysis\label{tab:dataset_summary}}
\tablehead{
\colhead{Sample} & 
\colhead{Selection Criteria} &
\colhead{Size} &
\colhead{Notes}
}
\startdata
\textbf{DESI DR1} & & & \\
Initial cross-match & APOGEE: \{\texttt{STARFLAG},\texttt{FE\_H\_FLAG},\texttt{MG\_FE\_FLAG}\}\texttt{=0}; SNR$>40$ &  &
 \\ 
 & DESI: \texttt{RVS\_WARN=0}; when available, \texttt{PRIMARY=True} & 8,104 &  \\
Training sample &
$T_{\rm eff}>4500$ K, 469 metal-poor + 1,600 metal-rich &
2,069 &
For fine-tuning or scratch training \\
Testing sample &
Remaining stars with $T_{\rm eff}>4000$ K &
5,539 &
For testing only \\
$[\alpha/\text{Fe}]$--[Fe/H] reference&
$T_{\rm eff}>4000$ K after initial cross-match&
7,608 &
For APOGEE [$\alpha$/Fe]--[Fe/H] reference\\
\textbf{DESI EDR} & & & \\[2pt]
Initial quality cuts &
\texttt{RVS\_WARN=0}, \texttt{PRIMARY=True}, \texttt{RR\_SPECTYPE=STAR} &
 & \\
 &
\texttt{MEDIAN\_COADD\_SNR\_B$>$20} &
134,137 &
 \\ 
Final EDR sample &
$T_{\rm eff}>4000$ K; valid [Fe/H],[$\alpha$/Fe]; good normalization &
126,253 &
For evaluating [$\alpha$/Fe]--[Fe/H] \\
\enddata
\end{deluxetable}

\section{Comparison with Clean Calibrated DESI SP Subset}
\label{sec:compare_clean}

In this section, we present the [Fe/H] and [$\alpha$/Fe] scatter plots obtained after applying a more restrictive set of quality cuts to the DESI SP sample. Following the quality cut process of \citet{2025arXiv250514787K} for comparisons between DESI and external catalogs, we impose the additional selections: \texttt{BESTGRID} $\neq$ \texttt{s\_rdesi1}, \texttt{RR\_SPECTYPE} = \texttt{STAR}, \texttt{SN\_R} $>$ 10, and \texttt{FEH\_ERR} $<$ 0.1, where \texttt{FEH\_ERR} is the [Fe/H] uncertainty reported by the DESI RVS pipeline.

\begin{figure*}
    \begin{center}
    \includegraphics[scale=0.25,angle=0]{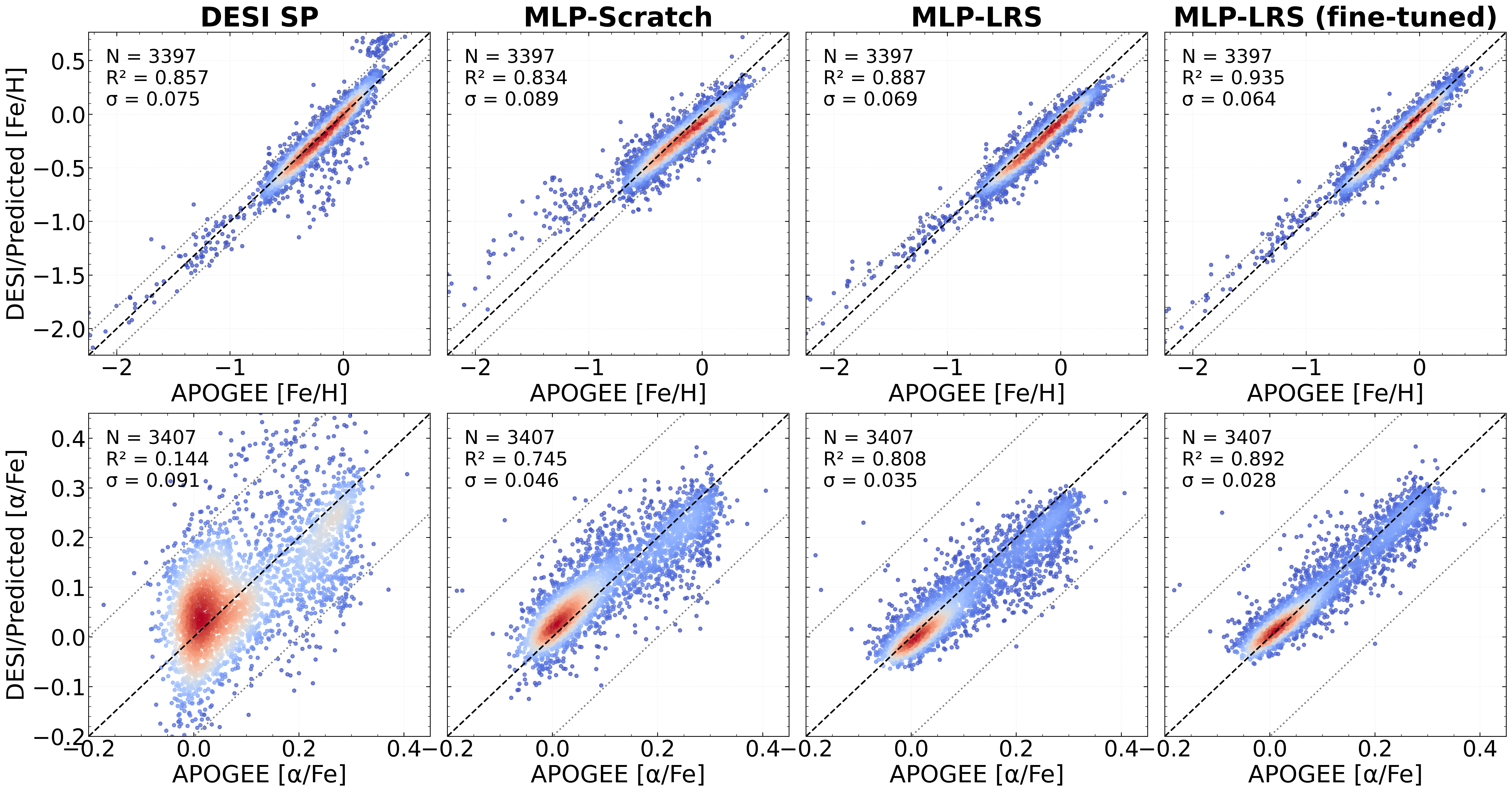} 
    \caption{Comparison of [Fe/H] and [$\alpha$/Fe] estimates between the clean, calibrated DESI SP subset and MLP-based models. Plotting conventions follow Figure~\ref{fig:compare_pretrained_desi}.}
    \label{fig:compare_pretrained_desi_clean}
    \end{center}
\end{figure*}

Following \citet{2025arXiv250514787K}, we further apply the recommended calibration to the DESI SP [Fe/H] values, noting that metallicity biases with respect to APOGEE and GALAH in the original DESI catalog show temperature-dependent trends that are well described by quadratic functions of $T_{\rm eff}$, with distinct relations for dwarfs and giants. The resulting comparisons are shown in Figure~\ref{fig:compare_pretrained_desi_clean}. Relative to Figure~\ref{fig:compare_pretrained_desi}, approximately 2,100 stars are removed by these stricter quality cuts. In addition, some stars have \texttt{[Fe/H]} reported as \texttt{NaN} after the calibration step and therefore do not appear in the scatter plot.

\begin{figure*}
    \begin{center}
    \includegraphics[scale=0.23,angle=0]{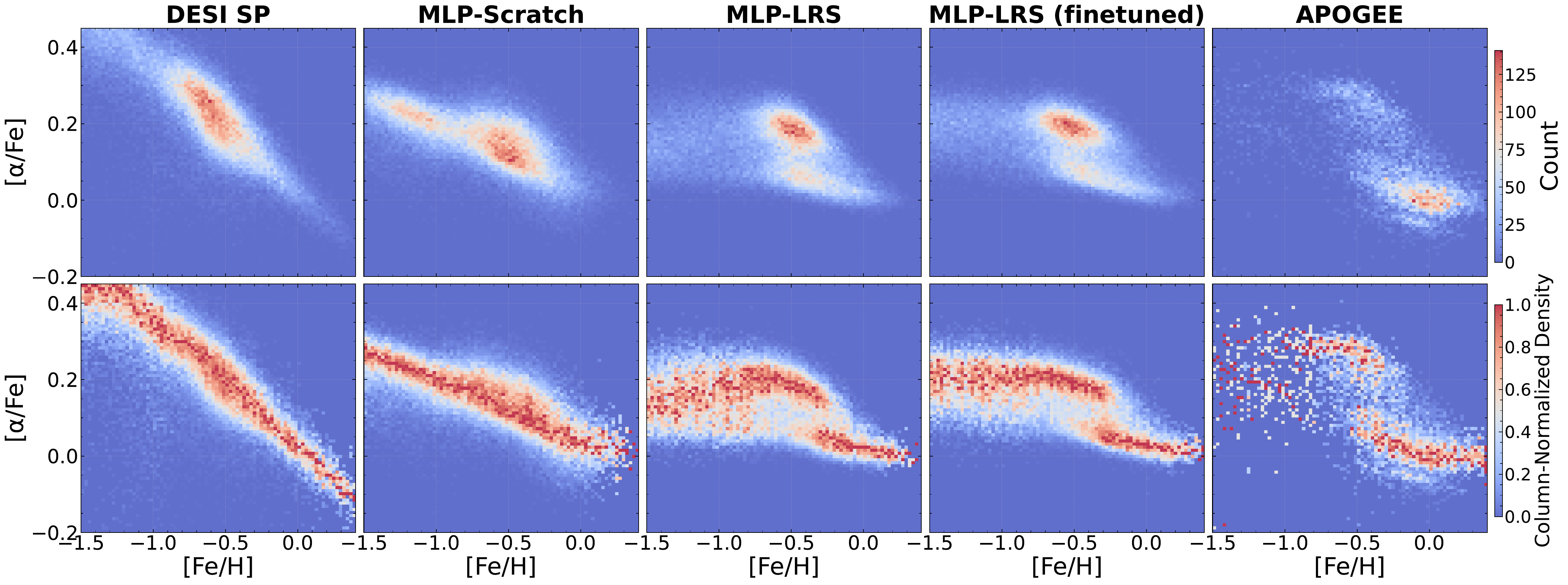} 
    \caption{[$\alpha$/Fe]–[Fe/H] diagrams for the DESI EDR and the APOGEE reference sample. A clean, calibrated DESI SP subset is used, with common sources shown in the second, third, and fourth columns. Plotting conventions follow Figure~\ref{fig:compare_pretrained_desi_afe_feh}.}
    \label{fig:compare_pretrained_desi_afe_feh_clean}
    \end{center}
\end{figure*}

For the [$\alpha$/Fe]–[Fe/H] diagrams, we also provide a version based on cross-matching the selected EDR sample from the main text with DR1 sources that have \texttt{PRIMARY=True}, in order to obtain the most up-to-date [Fe/H] and [$\alpha$/Fe] values (noting that EDR and DR1 use the same SP pipeline). We then apply the additional requirement \texttt{BESTGRID} $\neq$ \texttt{s\_rdesi1} and adopt the same [Fe/H] calibration as above. After these selections, 86,412 sources with valid stellar parameters remain from the originally selected EDR sample. The resulting comparison is shown in Figure~\ref{fig:compare_pretrained_desi_afe_feh_clean}.

\vfill\eject
\FloatBarrier
\bibliography{sppara_calib}{}
\bibliographystyle{aasjournalv7}
\end{document}